\documentclass[fleqn,usenatbib]{mnras}

\usepackage{newtxtext,newtxmath}

\usepackage[T1]{fontenc}

\DeclareRobustCommand{\VAN}[3]{#2}
\let\VANthebibliography\thebibliography
\def\thebibliography{\DeclareRobustCommand{\VAN}[3]{##3}\VANthebibliography}

\usepackage{graphicx}	
\usepackage{amsmath}	
\usepackage{tikz,xcolor,hyperref} 
\usepackage{gensymb}
\usepackage{tabularx,booktabs}
\usepackage{anyfontsize}
\usepackage{rotating}
\usepackage{xspace}     

\definecolor{lime}{HTML}{A6CE39}
\DeclareRobustCommand{\orcidicon}{%
    \begin{tikzpicture}
    \draw[lime, fill=lime] (0,0) 
    circle [radius=0.16] 
    node[white] {{\fontfamily{qag}\selectfont \tiny ID}};
    \draw[white, fill=white] (-0.0625,0.095) 
    circle [radius=0.007];
    \end{tikzpicture}
    \hspace{-2mm}
}

\newcommand{\orcidChrisO}{\href{https://orcid.org/0000-0003-0017-349X}{\orcidicon}}
\newcommand{\orcidChrisW}{\href{https://orcid.org/0000-0002-4569-016X}{\orcidicon}}

\newcommand{\orcidNA}{\href{https://orcid.org/0009-0005-7553-049X}{\orcidicon}}
\newcommand{\orcidRW}{\href{https://orcid.org/0000-0002-5325-2709}{\orcidicon}}
\newcommand{\orcidJH}{\href{https://orcid.org/0000-0001-8359-2328}{\orcidicon}}
\newcommand{\orcidSL}{\href{https://orcid.org/0000-0001-9372-4611}{\orcidicon}}

\newcommand{\refbf}{} 

\definecolor{asparagus}{rgb}{0.53, 0.66, 0.42}

\newcommand{\ha}{H$\alpha$\xspace}
\newcommand{\hb}{H$\beta$\xspace}
\newcommand{\hg}{H$\gamma$\xspace}

\newcommand{\oiii}{[O\,\textsc{iii}]\xspace}
\newcommand{\nii}{[N\,\textsc{ii}]\xspace}
\newcommand{\sii}{[S\,\textsc{ii}]\xspace}
\newcommand{\lha}{$L_{\mathrm{H}\alpha}$\xspace}
\newcommand{\lhb}{$L_{\mathrm{H}\beta}$\xspace}

\newcommand{\loiii}{$L_\mathrm{[OIII]}$\xspace}
\newcommand{\lcont}{$L_{5100}$\xspace}

\newcommand{\fwhmhb}{FWHM$_{\mathrm{H}\beta}$\xspace}
\newcommand{\fwhmha}{FWHM$_{\mathrm{H}\alpha}$\xspace}
\newcommand{\fwhmhg}{FWHM$_{\mathrm{H}\gamma}$\xspace}
\newcommand{\ratio}{H$\beta_\mathrm{Br}$/\oiii}

\title[The OzSSy1 Atlas]{OzSSy1: The Australian Southern Seyfert-1 Spectroscopic Atlas and Catalogue at $z<0.1$}

\author[Amrutha et al.]{
Neelesh Amrutha,$^{1}$\thanks{E-mail: neelesh.amrutha@anu.edu.au}\orcidNA
Christian Wolf,$^{1,2}$\orcidChrisW
Christopher A. Onken,$^{1,2}$\orcidChrisO
Wei Jeat Hon,$^{1,3}$\orcidJH
\newauthor{}
Samuel Lai,$^{4}$\orcidSL
and Rachel Webster$^{3}$\orcidRW
\\
$^{1}$Research School of Astronomy and Astrophysics (RSAA), Australian National University, Canberra ACT 2611, Australia\\
$^{2}$Centre for Gravitational Astrophysics (CGA), Australian National University, Building 38 Science Road, Acton ACT 2601, Australia \\
$^{3}$School of Physics, University of Melbourne, Parkville, Victoria 3010, Australia \\
$^{4}$Commonwealth Scientific and Industrial Research Organisation (CSIRO), Space \& Astronomy, P. O. Box 1130, Bentley, WA 6102, Australia\\
}

\date{Accepted XXX. Received YYY; in original form ZZZ}

\pubyear{\the\year{}}

\begin{document}
\label{firstpage}
\pagerange{\pageref{firstpage}--\pageref{lastpage}}
\maketitle

\begin{abstract}
We present a spectroscopic atlas of 1001 broad-line active galactic nuclei (AGNs) in the Southern sky, spanning redshifts $z < 0.1$, declinations $\delta <0\degree$ and Galactic latitudes $|b|>10\degree$.  The sample aims at being a comprehensive census of nearby broad-line AGNs at $K<13$ mag. The atlas is constructed from observations with $R\sim3000$ using the integral-field Wide Field Spectrograph at the Australian National University 2.3\,m telescope. The data products include spectra, each accompanied by a spectral decomposition that models the AGN continuum and host galaxy, along with fits to the emission lines \hg, \hb, \ha, He\,\textsc{ii}, He\,\textsc{i}, \oiii, [O\,\textsc{i}], \nii and [S\,\textsc{ii}], including broad Balmer and Helium components. The data products are publicly available and designed to support studies of population demographics and AGN variability in conjunction with future time-domain surveys. 
\end{abstract}




\section{Introduction}\label{sec:intro}

The spectral energy distribution of active galactic nuclei (AGNs) spans the entire electromagnetic spectrum \citep{brown_duncan_2019}. In the optical, emission from the viscous accretion disc and surrounding gas produces diagnostic emission lines whose luminosities, widths, ratios, and profiles encode the physical conditions of the broad- and narrow-line emitting regions (BLR and NLR). These features enable redshift measurements and single-epoch estimates of supermassive black hole (SMBH) masses, making large spectroscopic samples particularly valuable for AGN population studies.

Optical AGN samples are often incomplete at low luminosities and low redshifts, where host-galaxy starlight can dominate the observed emission and cause AGNs to be missed by photometric selection \citep{ho_filippenko_1997,ho_2008_rev,barquingonzalez_mateos_2024}. Spectroscopy is therefore necessary to unambiguously confirm the presence of an active nucleus in such systems, with sensitivity limited by the contrast between the point-source AGN and the extended host galaxy. This paper presents a spectroscopic atlas and catalogue of type 1 Seyfert galaxies in the Southern hemisphere at $z<0.1$, providing homogeneous spectral measurements for a defined sample of nearby broad-line AGN.

Seyfert galaxies \citep{seyfert_1943} are most readily identified at low redshifts, due to their intrinsically lower luminosity relative to quasars, enabling access to the longest rest-frame temporal baselines for characterising AGN variability. Accretion disc emission is intrinsically variable across timescales from days to decades, reflecting stochastic accretion processes and structural instabilities \citep{kelly_bechtold_2009,cackett_bentz_2021,arevalo_churazov_2024,blaes_jiang_2025}. Crucially, low-luminosity AGNs exhibit more rapid and higher-amplitude UV-optical variability than their more luminous quasar counterparts \citep{vanden_berk_wilhite_2004,tang_wolf_2023,tan_wolf_2026}, making nearby Seyferts preferred targets for reverberation mapping \citep[RM;][]{blandford_mckee_1982,peterson_ferrarese_2004} of the BLR \citep{wang_woo_2024} and other variability studies.

An increasing availability of wide-field time-domain photometry opens new windows on AGN variability. Current large photometric surveys include the Asteroid Terrestrial-impact Last Alert System \citep[ATLAS;][]{tonry_denneau_2018_atlas} and the Zwicky Transient Facility \citep[ZTF;][]{bellm_kulkarni_2019_ztf}. These all-sky or wide-area surveys have operated for nearly a decade and will be complemented and extended by the Legacy Survey of Space and Time \citep[LSST;][]{LSST_plan}, which will deliver unprecedented photometric cadence and depth over a ten-year baseline. Among the most striking phenomena uncovered are changing-look AGNs \citep[CLAGNs; see review by][]{ricci_trakhtenbrot_2022}, in which broad emission lines appear or disappear on timescales of months to years \citep{tohline_osterbrock_1976,macleod_ross_2016,macleod_green_2019,amrutha_wolf_2024}. Further varieties of nuclear transients are being discovered, including tidal disruption events \citep[TDEs;][]{gezari_tde_2021} and slow blue nuclear hypervariables \citep{lawrence_bruce_2016}. 

Interpreting these events requires comparison against a well-characterised archival reference spectrum \citep{lopez_martinez_2022,lopez-navas_sanchez-saez_2023,sanchez-saez_hernandez_2024_varOIII,fries_trump_2024}, but spectroscopic variability studies are limited by the expense of repeat observations. Large-scale programmes, such as the Sloan Digital Sky Survey Reverberation Mapping \citep[SDSS-RM;][]{shen_hall_2019} project, have monitored AGNs over multi-year baselines, but are restricted to small sky areas (e.g. a single $7\,\mathrm{deg}^2$ field) and modest sample sizes. Longer-baseline spectroscopic monitoring exists for only a handful of nearby AGN, including well-studied systems such as NGC~5548 and NGC~4151 \citep{bon_zucker_2016_ngc5548_lc,chen_bao_2023}, which have been observed for 43~years and 28~years respectively in observed frame. While these intensive campaigns have provided critical insights into AGN structure and variability, they are necessarily limited in their ability to generalise variability across the broader AGN population. Previous efforts to construct samples of AGN have largely relied on literature compilations of known quasars, such as \citet{veroncetty_veron_2010} and \citet{milliquas_2023}. These compilations inherit the heterogeneous selection criteria, and occasional misclassifications of the underlying surveys and catalogues, and do not provide a uniformly selected, homogeneous set of spectra for systematic analysis. The SDSS-V Black Hole Mapper \citep{kollmeier_rix_2026_SDSSV} currently represents the most ambitious ongoing effort; in the past, SDSS efforts have been concentrated in the Northern hemisphere, with the first large data release in the Southern hemisphere limited by the saturation limit for bright optical sources \citep{sdss_dr20}.

In the Southern hemisphere, the 6-degree Field Galaxy Survey \citep[6dFGS;][]{jones_saunders_2004_6dfgs,jones_read_2009_6dfgs} provides the currently most complete flux-limited spectroscopic catalogue of galaxies at $z<0.1$. Several studies have already exploited 6dFGS subsets as reference epochs for variability studies \citep{malyali_rau_2024,amrutha_wolf_2024,saha_markowitz_2025,wang_lin_2025,tan_wolf_2026}. However, the 6dFGS observations were obtained approximately two decades ago, a timescale comparable to the characteristic BLR orbital period in low-luminosity AGN \citep{amrutha_masscorr_2026}. Furthermore, the spectra are not flux-calibrated and were obtained using a wide-field Schmidt telescope with a small plate scale and a fibre diameter of $6.7$ arcsec, encompassing a substantial fraction of host-galaxy light for nearby AGN. While AGNs within 6dFGS have since been identified and their emission lines measured \citep{chen_zaw_2022,hon_webster_2025,suresh_hon_2025}, the original observations are not ideally suited as reference epochs for contemporary variability studies.

Re-observing the low-redshift 6dFGS AGN population therefore offers an opportunity to establish a modern, homogeneous spectroscopic reference sample with a multi-decade temporal baseline. Such a dataset enables population-level constraints on long-term spectroscopic variability, provides a benchmark for identifying anomalous behaviour and nuclear flaring events discovered in photometric surveys, serves as a pre-LSST spectroscopic reference ahead of future surveys such as the 4MOST Hemisphere Survey \citep{taylor_cluver_2023}, and complements the SDSS-V Black Hole Mapper spectra.

In this work, we present OzSSy1, the Australian Southern Seyfert-1 spectral atlas, a sample of 1001 AGNs at $z<0.1$ observed with the Wide Field Spectrograph (WiFeS) on the ANU 2.3-m telescope at Siding Spring Observatory. The data release comprises reduced integral field data cubes, extracted AGN spectra, along with AGN–host decomposition, and a catalogue of continuum and emission-line measurements, allowing a broad range of AGN population studies while providing a spectroscopic reference for variability studies. The sample is predominantly drawn from the 6dFGS broad emission line AGN sample by \citet{hon_webster_2025} but complemented with other selections of broad line AGNs. 

This paper is structured as follows: Section~\ref{sec:data} describes the target selection. Sections~\ref{sec:cubes} and \ref{sec:spectra} explain observations, data reduction, spectrum extraction and decomposition. Section~\ref{sec:validation} discusses the quality of the spectra, Section~\ref{sec:sample_props} the overall sample properties, and Section~\ref{sec:summary} summarises the paper. The data product structure is listed in Appendix~\ref{appendix_1}. Throughout the paper, we use AB magnitudes, with the exception of 2MASS-XSC magnitudes which are presented in Vega magnitudes, and adopt a flat $\Lambda$CDM cosmology with $\Omega_\Lambda = 0.7$ and $H_0 = 70$~km~s$^{-1}$~Mpc$^{-1}$.

\section{Sample definition}\label{sec:data}
\begin{figure*}
    \centering
    \includegraphics[width=0.93\linewidth]{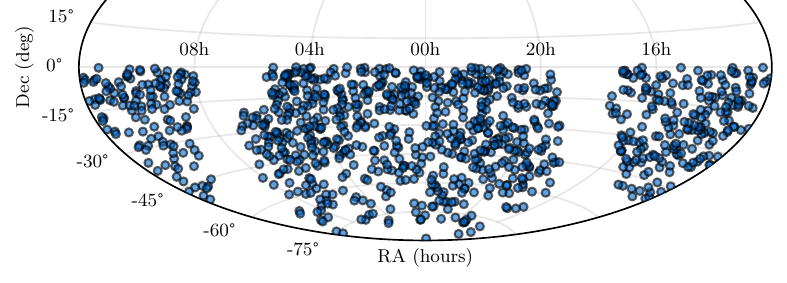}
    \caption{Distribution of the {\refbf 1001} AGNs in the atlas on the sky: The sample covers the Southern hemisphere at Galactic latitudes $|b| \geq 10\degree$.}
    \label{fig:sky_map}
\end{figure*}

This atlas aims to be as complete as possible for Southern broad-line AGNs at $z<0.1$, encompassing both previously catalogued objects and newly identified systems, subject to sufficient AGN-to-host contrast. The sample consists of 1001 AGNs, whose sky distribution is presented in Figure~\ref{fig:sky_map}. In the following, we explain the construction and completeness of the sample.

\subsection{Parent sample and target selection}

The majority of sources in our sample are drawn from the 6-degree Field Galaxy Survey \citep[6dFGS;][]{jones_saunders_2004_6dfgs,jones_read_2009_6dfgs}, a redshift survey with a primary target selection based on near-infrared (NIR) $K$-band magnitudes from the Two Micron All Sky Survey Extended Source Catalogue \citep[2MASS XSC;][]{jarrett_chester_2000}. The 6dFGS contains 125,071 galaxies complete to $K < 12.65$ mag, with a median redshift of $z \sim 0.053$, and covers the Southern sky at Galactic latitude $|b| \geq 10\degree$. 

From this parent galaxy sample, \citet{hon_webster_2025} identified a large sample of broad emission-line AGNs with Balmer emission line FWHM~$>1200$ km s$^{-1}$ including narrow-line Seyfert 1s, which is the first pillar of our target selection. We restrict this to $z < 0.1$, beyond which the \hb--\oiii complex approaches the strong $\lambda5577$\AA\ sky emission line, making accurate sky subtraction difficult. This redshift cut yields 784 broad-line AGN. We include all objects in this parent sample regardless of close foreground neighbours ($<3$ arcsec), bad 6dFGS spectra, or high Galactic extinction, since we independently re-observe the sample with a different instrument.

As a small supplement, we include low-redshift AGNs from the Siding Spring Southern Seyfert Spectroscopic Snapshot Survey \citep[S7;][]{thomas_dopita_2017}. These galaxies were originally selected to map ionisation cones in nearby ($z<0.023$) AGNs with high spatial resolution. While these observations were obtained with the same instrument (see Section~\ref{sec:cubes}), type 1 AGNs can be strongly variable, and we re-observe them to provide an updated spectroscopic dataset. Only eleven type 1-1.9 Seyferts were included, as type 2 objects are beyond the scope of this atlas.

Since 6dFGS is selected from the 2MASS XSC, it excludes point sources and is therefore incomplete for more luminous AGN, where emission dominates that of the host galaxy and produces a quasar-like appearance. To address this, we include AGN found by a sister project, the All-Sky Bright Complete Quasar Survey \citep[AllBRICQS;][]{onken_wolf_AllBRICQS_2023}. AllBRICQS is constructed from {\it Gaia} \citep[DR3;][]{gaia_dr3} and mid-infrared \textit{WISE} data \citep[Cat\textit{WISE}2020;][]{marocco_eisenhardt_catwise_2021} to identify bright quasars. The infrared selection begins with {\it Gaia} sources of $R_\mathrm{p} \leq 16$, Galactic latitude $|b| > 10\degree$, that are consistent with negligible parallax and proper motion. The AllBRICQS spectral atlas for quasars at $0.1<z<5$ will soon be published, while $z<0.1$ objects are presented here. AllBRICQS has provided 40 type 1 AGNs to this atlas, three from \citet{onken_wolf_AllBRICQS_2023}, and 37 new discoveries published here for the first time.

6dFGS was designed as a redshift survey, and $\sim$12 per cent of targets with existing archival spectroscopy were excluded from 6dFGS observations. Consequently, the 6dFGS AGN sample is not complete, motivating a supplementary selection from the Milliquas catalogue \citep{milliquas_2023}. Milliquas extends the compilation of known quasars presented by \citet{veroncetty_veron_2010} and incorporates objects identified up to 30 June 2023. Restricting Milliquas to the sky and redshift coverage of our sample ($z < 0.1$, $\delta < 0\degree$, and $|b| \geq 10\degree$) yields 3791 sources with type classifications containing either `A' or `Q', corresponding to type-1 AGN (host-dominated Seyferts) and quasars (core-dominated AGN), respectively. Although these classifications provide a high-confidence parent sample of active nuclei, the strength and visibility of broad emission lines are not uniformly characterised. We therefore further refine the Milliquas selection to favour sources for which broad-line emission is detectable in our observations. In particular, to limit the inclusion of host-dominated systems at low redshift, where broad-line signatures may fall below our sensitivity limit, we apply a redshift-dependent magnitude limit of $K < 13 + 3 \log(z/0.1)$ mag (Figure~\ref{fig:mag_z}), using \texttt{K\_m\_ext} from the 2MASS XSC.

After cross-matching with the 2MASS XSC within 15 arcsec and applying the magnitude selection, 1649 candidates remained. Of these, 1480 have a 6dFGS spectrum and were removed from the Milliquas candidate list because they had already been considered by \citet{hon_webster_2025} in a process that included visual vetting of spectra. The remaining 169 candidates were then assessed using all available spectroscopy. Our inclusion criterion is the presence of broad emission lines in at least one available spectrum, irrespective of the survey, observing epoch, or literature source. Conversely, a Milliquas type-1 classification alone is not sufficient for inclusion if the spectrum on which that classification is based can be inspected and does not show broad emission lines. If the referenced spectrum is unavailable, however, the object is retained in the sample. In addition to available literature spectra, we inspected spectra from Sloan Digital Sky Survey \citep[SDSS DR20;][]{sdss_dr20}, the BAT AGN Spectroscopic Survey \citep[BASS DR2;][]{oh_koss_2022}, and the Dark Energy Spectroscopic Instrument \citep[DESI DR1;][]{desi_dr1}. We also included known type-1 AGNs identified in these data releases that were not present in Milliquas (N$\sim$10), provided that broad emission lines were verified in their spectra.

This approach is particularly important for extremely variable AGNs, including those labelled as CLAGN, for which the optical spectral classification can depend strongly on the observing epoch. Therefore, rather than adopting a single catalogue classification, we considered the entire publicly available spectral record when determining whether an object satisfies the type-1 criterion. For the 1480 objects with 6dFGS spectra, for example, the majority are referenced by \citet{zaw_chen_2019} and \citet{chen_zaw_2022}, whose classifications were based on the 6dFGS spectra themselves. These spectra were inspected directly in \citet{hon_webster_2025}; objects for which broad emission was present were retained, whereas objects whose referenced spectra did not show broad lines were removed \citep[see discussion of these misclassifications in][]{hon_webster_2025}. The same principle was applied to the remaining Milliquas candidates in this work.

This procedure yielded a set of additional broad-line AGN that were not recovered through the 6dFGS, S7, or AllBRICQS selections and were subsequently incorporated into the sample. In total, 153 additional AGN were selected exclusively through Milliquas, the majority corresponding to objects included in the 6dFGS galaxy sample but lacking 6dF spectra. The original references associated with these Milliquas identifications are \citet{markaryan_lipovetskii_1973,bolton_shimmins_1975,deGrijp_lub_1987,mazzarella_boroson_1993,bowen_osmer_1994,sandage_bedke_1994,konig_staubert_1998,schwope_hasinger_2000,lamontagne_demers_2000,wisotzki_christlieb_2000,bauer_condon_2000,colless_dalton_2001,grazian_omizzolo_2002,paturel_petit_2003_pgc,sanders_mazzarella_2003,maia_machado_2003,crenshaw_kraemer_2003,bikmaev_sunyaev_2006,mauch_sadler_2007,kollatchny_kotulla_2008,huchra_macri_2012,monroe_prochaska_2016,liu_liu_2019_lozagn,zaw_chen_2019}.

After removing duplicate sources across all input catalogues, the final sample consists of 1001 AGNs. This combined dataset forms the basis of the IFU spectral atlas presented in this work. The heterogeneous origin of the sample naturally involves selection effects, particularly at the lowest AGN luminosities. These are discussed in the following section.

\subsection{Sample completeness}

\begin{figure*}
    \centering
    \includegraphics[width=1\linewidth]{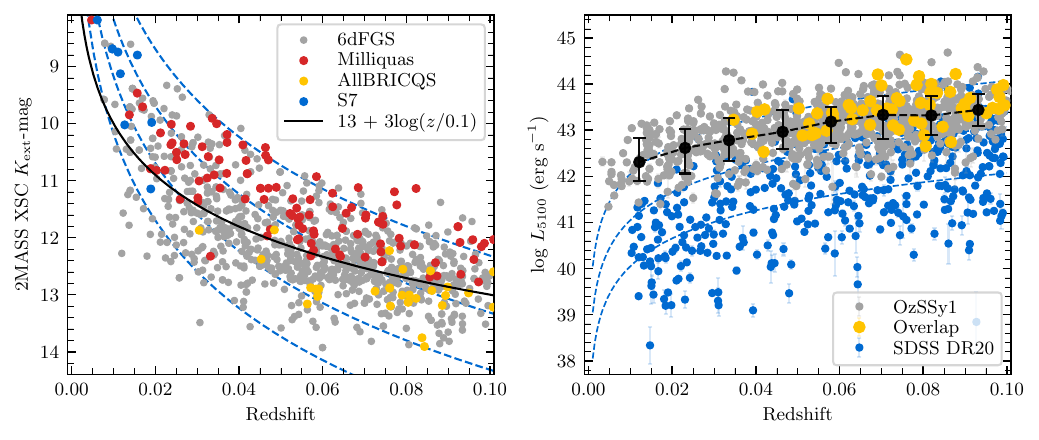}
    \caption{AGN sample. \textbf{Left:} $K$ magnitudes (\texttt{K\_m\_ext} column from 2MASS XSC) vs redshift. Dashed lines denote constant absolute magnitudes of $M_\mathrm{K}$: \{$-23$, $-24$, $-25$, $-26$\}. The solid line marks the heuristic $K < 13 + 3\log(z/0.1)$ limit we use for the Milliquas selection. The source surveys of the AGNs are marked separately. \textbf{Right:} The AGN monochromatic luminosity of the continuum, \lcont ($\lambda L_\lambda$ at $\lambda5100$\AA), as a function of redshift. The running median for the OzSSy1 sample is shown with large markers, with error bars indicating the 16th–84th percentile range. Dashed lines denote constant apparent flux (erg s$^{-1}$ cm$^{-2}$ \AA$^{-1}$): \{10$^{-15}$, 10$^{-16}$, 10$^{-17}$\}. Objects from the SDSS DR20 Quasar Spectra Property Catalog and the subset that overlaps with OzSSy1 are marked separately.}
    \label{fig:mag_z}
\end{figure*}

The parent 6dFGS sample is complete to $K < 12.65$ mag, brighter than the completeness limit of the 2MASS XSC ($>90$ per cent complete at $K \sim 13.57$ mag; \citealt{mcintosh_bell_2006}). Since 6dFGS is a galaxy survey rather than an AGN-targeted survey, the spectroscopic identification of type 1 AGNs by \citet{hon_webster_2025} ensures that the completeness of the 6dFGS AGN sample inherits the completeness of the underlying galaxy population at this magnitude limit, with the exception of the 12 per cent of the galaxies in the 6dFGS catalogue without 6dF spectra. Assuming the AGN fraction within this 12 per cent is the same as the entire sample, we expect 890 (784/0.88) AGNs in the full 6dFGS catalogue, or 106 additional AGNs to be added to the atlas. The Milliquas supplement recovers $\sim80$ per cent of these AGNs without 6dF spectra and extends coverage to point-source-like quasars absent from the 2MASS XSC. Therefore, the AGN sample in this atlas has at least inherited the galaxy completeness of 6dFGS and 2MASS XSC for extended sources.

A limitation for the sample completeness arises from AGN variability. The 2MASS photometry, obtained more than two decades prior to the WiFeS observations, may not reflect the AGN luminosity at the epoch of spectroscopic follow-up. AGN luminosities can vary by factors of $\sim$2 on average over such timescales \citep{amrutha_masscorr_2026,tan_wolf_2026}, causing sources to scatter across the adopted magnitude boundary in either direction. All type-2 AGNs in the 6dFGS catalogue \citep{suresh_hon_2025} exhibiting photometric variability have been followed up spectroscopically to identify the emergence of broad emission lines \citep[i.e., turn-on CLAGN;][]{hon_wolf_2022_skymapper_colors,amrutha_wolf_2024}, where variability is quantified using 7-yr ATLAS light curves in the orange band and defined by the variability amplitude relative to a non-variable star-forming galaxy comparison sample. Turn-off CLAGN are already included through their earlier broad-line classifications in 6dFGS.

Other compilations of CLAGN \citep[see list in][ and references therein]{camus_panda_2026} are largely drawn from Northern-hemisphere surveys, extend beyond $z \sim 0.1$, or are constructed from previously known type-1 AGN samples, favouring the identification of turn-off events. Furthermore, most CLAGN classifications are based on changes in the broad \hb line, whereas broad \ha is often present in both spectral states and would therefore already satisfy our selection criteria. These considerations indicate that the present sample captures the vast majority of CLAGN within the defined selection boundary, with the primary remaining incompleteness arising from short-lived on-off transitions in type-2 AGNs not observed spectroscopically during their active phase.

The redshift-dependent magnitude limit, $K < 13 + 3\times\log(z/0.1)$, was introduced as a pragmatic selection criterion for the Milliquas supplement. At low redshift and luminosity, the host galaxy increasingly dominates the observed $K$-band emission, reducing the AGN-to-host contrast and consequently the detectability of broad emission lines. The adopted redshift dependence therefore provides a conservative way of restricting the Milliquas candidates to systems that are sufficiently luminous in the near-infrared to be more likely to have a detectable nuclear contribution, while avoiding a large population of nearby, host-dominated systems. We emphasise that this boundary is heuristic rather than being derived from a quantitative, but necessarily simplistic model, of the AGN-to-host contrast, and is therefore not intended to represent a physical completeness limit. Rather, it provides a practical compromise between retaining potentially relevant broad-line AGN and limiting contamination by sources for which the nuclear emission is unlikely to be detectable in the available spectroscopy.

Importantly, this criterion is applied only to the Milliquas-derived candidate sample, which is used to supplement the primary 6dFGS-based selection. The S7 and AllBRICQS samples provide additional complementary pathways for recovering broad-line AGN that may be absent from 6dFGS or from the literature compilation represented by Milliquas, and these sources are not subjected to the same $K$-band selection. This is particularly relevant for AllBRICQS, which targets the brightest sources in the sky using \textit{Gaia} $R_{\rm p}$ magnitudes rather than near-infrared galaxy luminosity. The combination of these independent selection pathways therefore reduces the sensitivity of the final sample to the precise choice of the Milliquas $K$-band boundary.

As an additional completeness check, we compared the OzSSy1 sample with the SDSS DR20 Quasar Spectral Properties Catalogue \citep{sdss_dr20}. DR20 is the first major SDSS quasar catalogue released since Milliquas v8 and, importantly, the first SDSS quasar catalogue to include a substantial number of southern sources. Of the 540 catalogue sources satisfying $z<0.1$, $\delta<0\degree$, and $|b|>10\degree$, 63 match OzSSy1 within 15 arcsec, all with separations below 1.3 arcsec, with the limited overlap reflecting the complementary selection of the two samples. To determine whether SDSS DR20 type-1 AGNs absent from OzSSy1 satisfy the OzSSy1 selection, we cross-matched them with Vista Hemisphere Survey \citep[VHS DR5;][]{mcmahon_banerji_vhs_2013,vhs_dr5_cat_2021}. None of the 307 sources with available aperture-corrected $K_S$ magnitudes satisfy the adopted $K < 13 + 3\log(z/0.1)$ criterion, and only six lie within one magnitude of the threshold. Despite differences between the VHS and 2MASS photometric measurements, these results suggest that few, if any, unmatched SDSS sources are excluded solely by the $K$--$z$ selection. In Figure~\ref{fig:mag_z} we show that the SDSS DR20 sources overlapping OzSSy1 occupy the high-luminosity end of the SDSS distribution, while the lower luminosity boundary of OzSSy1 is set primarily by the integrated $K$-band host-galaxy magnitude. Consequently, OzSSy1 includes nearby systems with relatively low-luminosity broad-line AGN hosted by massive galaxies.

We therefore regard the OzSSy1 sample as largely complete for type-1 AGN satisfying $z<0.1$ and $K < 13 + 3\times\log(z/0.1)$ within the limitations of broad-line detectability against host galaxy backgrounds in ground-based seeing. For the broader range $K<13$ mag, the sample remains substantially complete, although some residual incompleteness is expected where broad-line emission is strongly diluted by the host-galaxy continuum. This limitation is most important at low redshift ($z\lesssim0.03$), where the host galaxy contributes substantially to the integrated $K$-band flux, making the distinction between host-dominated and AGN-dominated systems increasingly ambiguous. A secondary source of incompleteness arises from AGN variability, which can move objects across the effective magnitude boundary between the epoch of the 2MASS photometry and the spectroscopic observations. The quoted completeness should therefore be interpreted as completeness within the empirically defined selection regime, rather than as a formal completeness measurement over the full type-1 AGN population.

\section{Data}\label{sec:cubes}

\subsection{Observations}

Observations were carried out at the Australian National University (ANU) 2.3-m telescope \citep{mathewson_hart_2013} at Siding Spring Observatory. We used the Wide Field Spectrograph \citep[WiFeS;][]{dopita_hart_2007,dopita_rhee_2010}, an integral field unit (IFU) with a field of view of $38\times25$ arcsec$^2$ sampled by $1$ arcsec$^2$ spaxels, corresponding to spatial scales of 200 pc at $z = 0.01$ and 1.9 kpc at $z=0.1$. We adopt the B3000 and R3000 gratings, providing a combined wavelength coverage of $3200$--$9800$~\AA\ with a mean resolution of $R\simeq3000$, corresponding to a velocity resolution of $\sim100$ km s$^{-1}$.

The observations for this sample were obtained between 2020 and 2026 (MJD 59015--61147), with approximately half of the spectra acquired during the 2023 calendar year. The observing campaign expanded significantly following the transition of the ANU 2.3-m telescope to robotic operations in March 2023 \citep{price_nielsen_robotic_2024}. Spectra obtained prior to this date were drawn from archival observations conducted for other programmes, including those presented in \citet{hon_wolf_2022_skymapper_colors} and \citet{amrutha_wolf_2024} for the identification of Changing-Look AGN.

Individual targets were typically observed in a single exposure with integration times ranging from 350 to 1800\,s. Exposure times were selected based on the $r$-band magnitudes from the SkyMapper Southern Sky survey \citep[SMSS DR4;][]{wolf_onken_2018_smss,onken_wolf_2024_smssdr4}. Observations were conducted during dark or grey time ($<50$ per cent lunar illumination), with the brightest targets observed under grey conditions. In some cases, observations were made at higher lunar illuminations, with large angular separation from the Moon. The average seeing conditions at the observatory is 2 arcsec.

While preparing robotic observation blocks, the instrument field of view was occasionally shifted by a few arcseconds or rotated to avoid bright neighbouring sources entering the IFU field. For extended targets, we positioned the field such that a region free of host-galaxy emission was available for sky estimation. In cases where the host galaxy extended across the entire field of view, we instead used the nod-and-shuffle observing mode. In this configuration the telescope alternates between the target and a nearby blank sky position, allowing the sky background to be measured separately. Although this approach effectively doubles the required observing time, these objects are typically the brightest and nod-and-shuffle mode provides significantly improved sky subtraction compared to the standard observing mode.

\subsection{Data cube generation}

The integral field data cubes were reduced using an updated version (Onken at al., in prep) of the \texttt{PyWiFeS} pipeline described by \citet{childress_vogt_2014_pywifes}. We briefly summarise the reduction procedure here. The \texttt{PyWiFeS} pipeline performs standard CCD reduction and cube reconstruction procedures. Raw frames are first bias-subtracted and corrected using dome flat-field exposures to remove pixel-to-pixel sensitivity variations. Wavelength calibration is performed using arc-lamp exposures, producing a wavelength solution for each slitlet. The data are then rectified and assembled into a three-dimensional data cube with two spatial dimensions and one spectral dimension. The cubes are corrected for atmospheric differential refraction, which causes an excess deflection of light of bluer wavelengths. Flux calibration is implemented using observations of spectrophotometric standard stars with identical instrument settings and extracting their instrumental spectra in a circular aperture with 12~arcsec diameter. Telluric absorption features are corrected using the same standard star observations for just the red arm. For observations conducted in nod-and-shuffle mode, sky subtraction is performed directly within the pipeline using the interleaved sky exposures, while for standard observations the sky subtraction is handled during the spectral extraction stage as described in Section~\ref{sec:spectra}. The final products of the pipeline are wavelength-calibrated and flux-calibrated integral field data cubes for both the blue and red spectrograph arms. The two arms are independently reduced, and the cubes are re-binned in the wavelength axis, with even bin widths of 0.77\AA\ for the blue arm and 1.25\AA\ for the red arm.

\subsection{Spectrum extraction}\label{subsec:spec_extr}

We extract integrated AGN spectra from the reduced IFU data cubes after subtracting sky emission. For observations obtained in nod-and-shuffle mode, sky subtraction is performed during cube generation. For the remaining observations, we estimate the sky spectrum directly from the cube by selecting a circular region with a diameter of 5 arcsec and subtracting its median spectrum from the entire cube. This aperture balances reliable sky estimation against spatial variations in the background. For WiFeS data, where each vertical strip in the spatial plane shares consistent wavelength calibration, we select the sky region along the same spatial column as the AGN centroid wherever possible to minimise residuals around prominent sky emission lines.

\begin{figure*}
    \centering
    \includegraphics[width=1\linewidth]{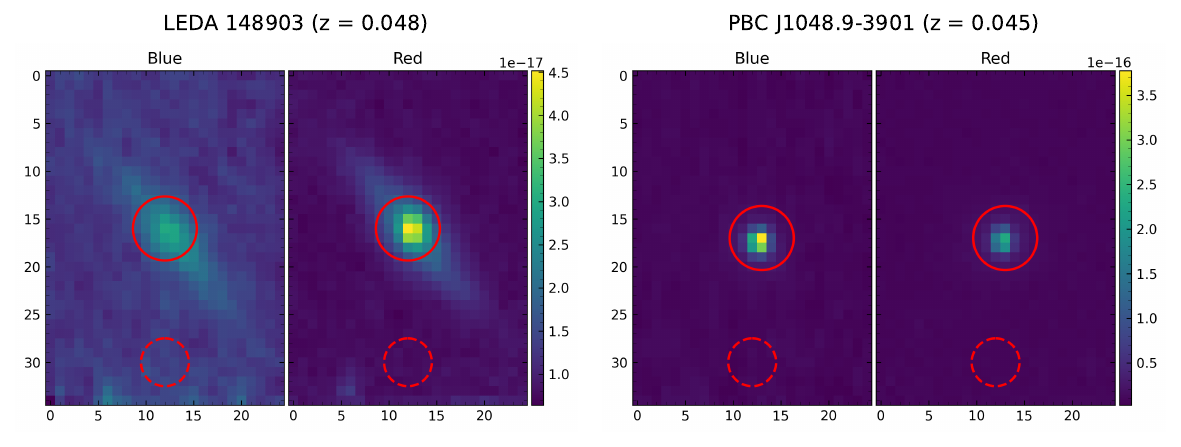}
    \caption{Example spatial extraction images of two sources at a similar redshift. \textbf{Left:} Source with visible extended host-galaxy. \textbf{Right:} Source that is AGN-dominated. For each object, the WiFeS data cube is collapsed into median images in the blue and red arms, with flux units of erg s$^{-1}$ cm$^{-2}$ \AA$^{-1}$. The solid circle indicates the 6.7 arcsec extraction aperture, and the dashed circle marks the 5 arcsec sky region used for background subtraction.}
    \label{fig:spat_grid}
\end{figure*}

Given the low redshifts of the sample and the 1 arcsec$^2$ spatial resolution of WiFeS, most host galaxies are spatially resolved in the IFU data, making annular sky subtraction unsuitable due to their diverse morphologies and spatial extents. Instead, the AGN centroid and sky region were selected manually using \texttt{PySpecExtract} \citep{pyspecextract_2026_v2}. This allowed reliable identification of the nuclear source in crowded fields and ensured that sky regions were placed in areas free from host galaxy contamination. The extraction aperture centroid was then refined using a two-dimensional Moffat fit to the spatial profile, initialised at the manually identified nuclear position. Although we explored a wavelength-dependent PSF-optimised extraction algorithm, we adopt a fixed circular aperture to avoid wavelength-dependent changes in the extracted host contribution arising from variations in seeing, instrumental defocus and host-galaxy contamination of the PSF. The extracted spectrum is obtained by summing the spaxels within this aperture after sky subtraction. Example images of the extraction are provided in Figure~\ref{fig:spat_grid}.

\begin{figure}
    \centering
    \includegraphics[width=1\linewidth]{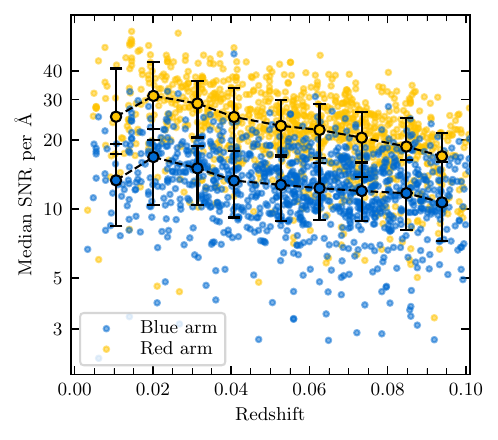}
    \caption{Distribution of the signal-to-noise ratio of the spectra, computed by taking the median signal to noise ratio of the extracted spectrum in the wavelength axis, as a function of redshift. The running medians for the red and blue arms are shown separately, with error bars indicating the 16th–84th percentile range.}
    \label{fig:snr_z}
\end{figure}

The spectra extracted for this atlas span the wavelength range $3800$--$9000$~\AA\ in observed frame. Instrumental sensitivity decreases blueward of $\sim4100$~\AA, while imperfect sky subtraction affects wavelengths redder than $\sim7800$~\AA. Nevertheless, this spectral range comfortably includes the Balmer emission lines from \ha to \hg; the latter may be difficult to detect in some cases due to the increased noise at the blue end and its own intrinsic weakness. The sample has mean signal-to-noise ratios (SNRs) of 13 and 23 per \AA\ in the blue and red arms, respectively (Figure~\ref{fig:snr_z}), where the SNR is defined as the median flux-to-error ratio across the extracted wavelength range.

\begin{figure}
    \centering
    \includegraphics[width=1\linewidth]{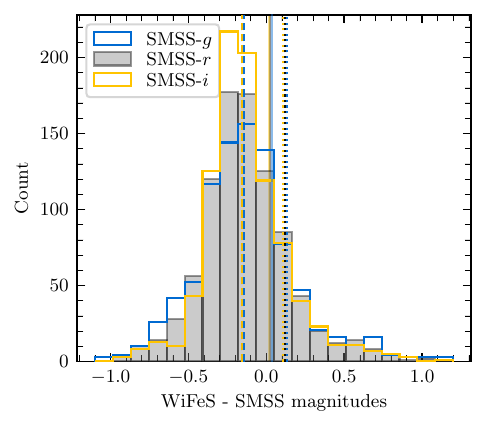}
    \caption{Distribution of differences in magnitudes extracted from the 6.7 arcsec aperture WiFeS spectra, with 6.0 arcsec magnitudes from SMSS DR4 in $g$, $r$ and $i$ bands. The dashed lines indicate medians of the differences. The dotted lines indicate the medians of the differences for 8.0 arcsec SMSS photometry. The solid lines indicate a linear interpolation to 6.7 arcsec aperture magnitudes.}
    \label{fig:mag_diff}
\end{figure}

The WiFeS blue and red arms are combined at $5650$~\AA, with the red arm scaled to match the blue arm to correct for small flux discontinuities at the transition. This correction is guided by comparisons with standard-star calibrations, and a $\sim100$~\AA\ overlap region centred at separation wavelength. The applied scaling factor is $0.96 \pm 0.12$ (mean and standard deviation), indicating a small systematic overestimate of the red-arm flux. The origin of this systematic offset is not currently understood. We assess the overall flux calibration by comparing synthetic $gri$ magnitudes from the extracted WiFeS spectra with aperture photometry from SMSS DR4 (Figure~\ref{fig:mag_diff}). After accounting for the difference between the WiFeS and SMSS aperture sizes, the median magnitude offsets are consistent with zero to within $\sim$0.04 mag, indicating good overall flux calibration. The scatter of $0.32$--$0.37$ mag across the three bands is consistent with intrinsic AGN variability over the $\sim$6-year interval between the SMSS and WiFeS observations, based on photometric variability analysis conducted by \citet{tan_wolf_2026}. Consistent with expectations, the scatter is larger in $g$-band than in $i$-band, reflecting the stronger variability of AGN emission as well as smaller average host fraction at shorter wavelengths.

\section{Spectral analysis}\label{sec:spectra}
\subsection{Spectrum decomposition}\label{sec:decomp}

\begin{figure*}
    \centering
    \includegraphics[width=0.99\linewidth]{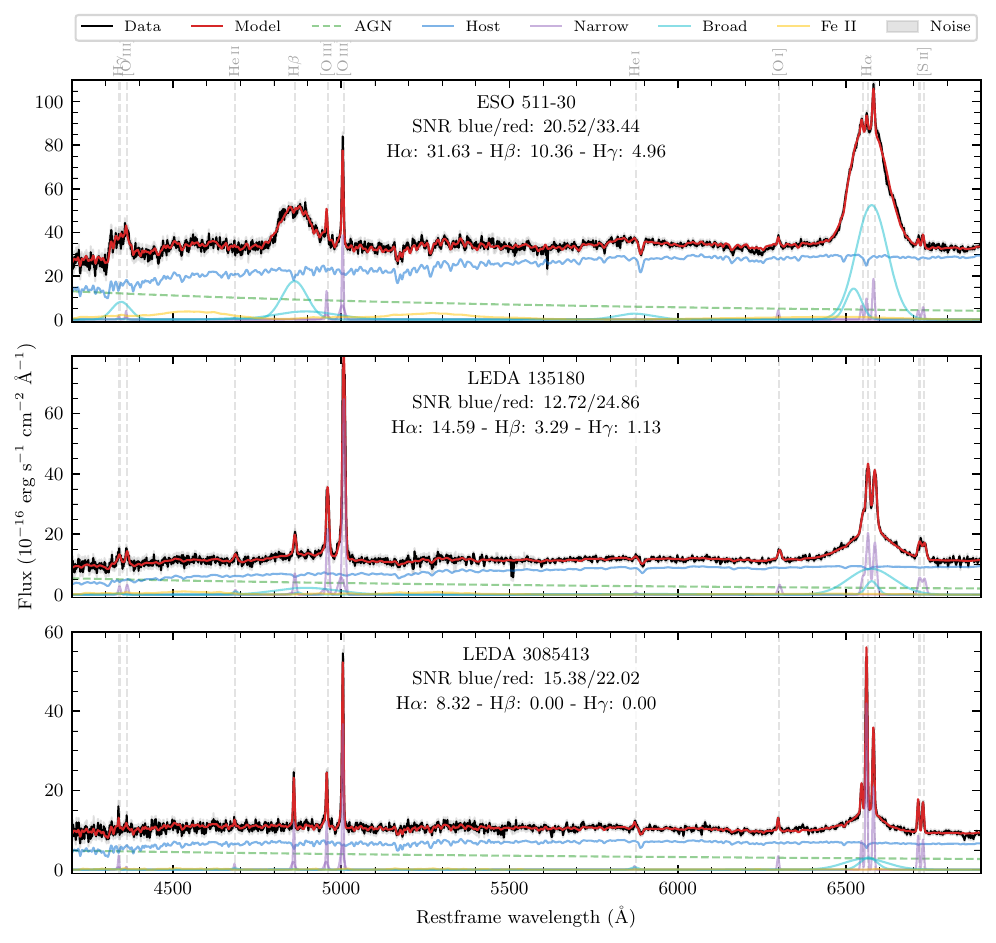}
    \caption{Spectrum decomposition: Examples with diverse \ratio ratios and Balmer line SNRs, representative of average AGNs in the sample. Vertical dashed lines denote centroids of emission lines based on Table~\ref{tab:line_list}. The median SNR per \AA\ of the two arms and the Balmer emission line SNRs are provided.}
    \label{fig:spec_decomp}
\end{figure*}

\begin{table*}
\centering
\caption{Emission line components used in the spectral fitting. Narrow components are tied to the kinematics of \oiii where indicated, while broad components are fitted independently. Vacuum wavelengths are used for line centroids, based on WiFeS wavelength calibration, and wavelengths refer to rest frame values.}\label{tab:line_list}
\begin{tabular}{lccc p{12.5cm}}
\toprule
\textbf{Line} & $\lambda_0$ (\AA) & N$_{\rm narrow}$ & N$_{\rm broad}$ & \textbf{Notes / Constraints} \\
\midrule

\hg      & 4341.68 & 2 & 2 & Narrow components tied to \oiii $\lambda5007$ core and wing kinematics and core-to-wing amplitude ratio. All broad component parameters are free. \\

{\oiii}      & 4364.44 & 2 & 0 & Narrow components tied to \oiii $\lambda5007$ core and wing kinematics and core-to-wing amplitude ratio. \\

He\,\textsc{ii}   & 4686.00 & 1 & 1 & Narrow component tied to \oiii $\lambda5007$ core component kinematics. Broad component shares velocity offset with \oiii $\lambda5007$ core component but otherwise free. \\

\hb       & 4862.69 & 2 & 2 & Narrow components tied to \oiii $\lambda5007$ core and wing kinematics and core-to-wing amplitude ratio. All broad component parameters are free. \\

{\oiii}     & 4960.30 & 2 & 0 & Flux fixed to 1/2.98 of \oiii $\lambda5007$; Core and wing kinematics tied to corresponding \oiii $\lambda5007$ components. \\

{\oiii}     & 5008.24 & 2 & 0 & Core and wing components fitted freely. Core component defines reference kinematics for all narrow-lines fit with one narrow component, while both core and wing components are used for narrow-lines fit with two narrow components. Wing component weaker and broader than core component. \\

He\,\textsc{i}     & 5875.61 & 1 & 1 & Narrow component tied to \oiii $\lambda5007$ kinematics. Broad component shares velocity offset with \oiii $\lambda5007$ core component but otherwise free. \\

{[O\,\textsc{i}]}   & 6300.30 & 1 & 0 & Narrow component tied to \oiii $\lambda5007$ core component dispersion and \nii $\lambda6585$ velocity offset. \\

{\nii}      & 6549.86 & 1 & 0 & Flux fixed to 1/2.93 of \nii $\lambda6585$; kinematics tied to \oiii $\lambda5007$ core component dispersion and \nii $\lambda6585$ velocity offset. \\

\ha      & 6564.63 & 1 & 2 & Narrow component tied to \oiii $\lambda5007$ core component dispersion and \nii $\lambda6585$ velocity offset. All broad component parameters are free. \\

{\nii}      & 6585.28 & 1 & 0 & Narrow component fitted with free amplitude and velocity offset; dispersion tied to \oiii $\lambda5007$ core component. \\

{\sii}      & 6718.29 & 1 & 0 & Narrow component tied to \oiii $\lambda5007$ core component dispersion and \nii $\lambda6585$ velocity offset. \\

{\sii}      & 6732.67 & 1 & 0 & Narrow component tied to \oiii $\lambda5007$ core component dispersion and \nii $\lambda6585$ velocity offset. \\

\bottomrule
\end{tabular}
\end{table*}

The extracted spectra are decomposed with the Bayesian AGN Decomposition Analysis for SDSS Spectra code \citep[BADASS3;][]{sexton_matzko_2021_badass3}. This framework models the full optical spectrum simultaneously, with the continuum, host galaxy contribution, and emission-line components fitted in a self-consistent manner. Unlike traditional approaches that require predefined continuum windows or manual masking of emission lines, BADASS3 fits the continuum and emission-line components jointly, reducing user intervention and enabling efficient modelling of large spectroscopic samples.

Each spectrum is fitted over the rest-frame wavelength range $4200$--$7100$~\AA, after correcting for Galactic extinction \citep{dust_map_SandF_2011}. The continuum model consists of three primary components: (i) a power-law representing the AGN continuum emission, (ii) an Fe\,\textsc{ii} pseudo-continuum based on the empirical template of \citet{veron_cetty_2004_feii_temp}, and (iii) a host galaxy component. Host-galaxy is modelled using a combination of single stellar population templates from the EMILES library \citep{vazdekis_2016_emiles_host}. The template basis is extended with a reddening parameter to account for intrinsic host-galaxy extinction (see Section~\ref{sec:cont_sci}). This flexibility is required for strongly reddened systems, such as highly inclined galaxies where part of the bulge is observed through dust lanes. All components are fitted simultaneously along with emission lines within a Bayesian framework, with parameter uncertainties derived from the posterior distributions.

Emission lines are represented using Gaussian components. Narrow emission lines are modelled with up to two Gaussian components in order to account for potential asymmetries or extended wings commonly observed in AGN narrow-line regions. Broad-line emission is modelled with two additional Gaussian components where present. The set of emission lines included in the fits, together with the adopted number of components for each line, is listed in Table~\ref{tab:line_list}. Vacuum wavelengths are used for line centroids, and the wavelengths refer to rest frame values.

To reduce degeneracy between the narrow and broad emission-line components, we use the kinematics of \oiii $\lambda5007$ as a reference for all narrow emission lines. The \oiii profile is modelled using two components: a core component representing the bulk of the NLR emission, and a wing component intended to capture outflowing gas \citep{sexton_matzko_2021_badass3}. These core and wing components are simultaneously applied to \oiii $\lambda4959$, narrow \hb, and narrow \hg, while all remaining narrow lines are modelled using only the core component. This approach is motivated by the fact that \oiii $\lambda5007$ is typically the strongest and least blended narrow emission line in the optical spectrum, providing the most reliable constraint on NLR kinematics. Adopting its line profile for narrow \hb and \hg reduces the degeneracy between the narrow and broad Balmer components and improves the stability of the decomposition.

In contrast, we model the narrow \ha and \nii emission using only a single component. The strong blending within the \ha-\nii complex causes the wing component to become highly degenerate with the broad \ha profile, even when its kinematics are tied to those of \oiii. Furthermore, the kinematics of \oiii are not always representative of those of the lower-ionisation narrow-line gas traced by \nii and \sii \citep{kovacevic_dojcinovic_2022}. Allowing a second narrow component in the \ha-\nii complex therefore introduces additional degeneracies without improving the constraints, and was found to degrade the reliability of the broad-line measurements.

The velocity dispersion of individual narrow components is restricted to $\sigma < 500$~km~s$^{-1}$, while broad components are required to satisfy $\sigma > 500$~km~s$^{-1}$, with an upper limit of $5500$~km~s$^{-1}$ for broad components and a lower limit of $40$~km~s$^{-1}$ for narrow components. Broad line components therefore have a FWHM range of $\sim 1200$~km~s$^{-1}$ to $14000$~km~s$^{-1}$, and narrows lines have a lower FWHM limit of $\sim 100$~km~s$^{-1}$, which corresponds to the resolution limit of WiFeS. In about 10 per cent of the objects, the broad line dispersion/FWHM were increased to $7500$~km~s$^{-1}$/$19000$~km~s$^{-1}$ to account for special cases, although the limits were not reached for majority of the fits. Example decompositions are shown in Figure \ref{fig:spec_decomp}.

\subsection{Spectral measurements in the catalogue}

\begin{table*}
\centering
\caption{Summary of extracted spectral fitting parameters, descriptions, and units. Lower and upper errors, when subtracted/added to the scalar, provide the 16th to 84th percentile range of the posterior distribution of the MCMC sampling. Wavelengths refer to rest frame values. Emission line properties are provided for the summed components where more than one component is used.}\label{tab:extracted_pars}
\begin{tabular}{lccl}
\toprule
\textbf{Parameter} & \textbf{Units} & \textbf{Type} & \textbf{Description} \\
\midrule

\multicolumn{4}{c}{\textbf{Basic Properties}} \\\hline
spectrum\_id      & -- & ID & Spectrum identifier \\
name, alt\_name   & -- & ID & Object names \\
z           & --  & scalar & Redshift \\
ra, dec     & deg & scalar & Right ascension, Declination \\
mjd         & day & scalar & Modified Julian Date \\
airmass     & --  & scalar & Airmass at the time of observation \\
ebv         & mag & scalar & E($B-V$) reddening value from \citet{dust_map_SandF_2011} \\
type        & --  & -- & Seyfert classification described in Section~\ref{subsec:sytype} \\
smss\_id    & --  & ID & Crossmatched SMSS DR4 ID of the object\\
smss\_dist & arcsec & scalar & On-sky separation of the source to crossmatched object\\
closest\_smss\_id, closest\_gaia\_id & -- & ID & Closest SMSS and \textit{Gaia} objects ($<$15 arcsec)\\
closest\_smss\_dist, closest\_gaia\_dist & arcsec & scalar & On-sky separation of the source to closest SMSS and \textit{Gaia} objects \\\toprule
\multicolumn{4}{c}{\textbf{Magnitudes}} \\\hline
r\_psf, e\_r\_psf & mag & scalar & SMSS $r$-band magnitudes used for exposure time calculation\\ 
spec\_\{g,r,i\}\_mag, spec\_\{g,r,i\}\_err & mag & scalar & 6.7 arcsec aperture magnitudes from WiFeS based on SMSS filters\\ 
smss\_\{g,r,i\}\_magc06, smss\_\{g,r,i\}\_errc06 & mag & scalar & Mean seeing corrected 6.0 arcsec aperture magnitudes from SMSS DR4 \\ 
\{j,h,k\}\_mag\_ext, \{j,h,k\}\_mag\_err & mag & scalar & 2MASS XSC magnitudes \\ 
\toprule

\multicolumn{4}{c}{\textbf{Emission Line Properties}} \\\hline
*\_lum, *\_lum\_low / upp   & log erg s$^{-1}$ & scalar & Emission line luminosity, lower / upper uncertainty on luminosity\\
*\_fwhm, *\_fwhm\_low / upp & log km s$^{-1}$ & scalar & Full width at half maximum, lower / upper uncertainty on FWHM \\
*\_disp, *\_disp\_low / upp & log km s$^{-1}$ & scalar & Velocity dispersion, lower / upper uncertainty on dispersion \\
*\_ew, *\_ew\_low / upp & \AA & scalar & Equivalent width, lower / upper uncertainty on EW \\\toprule

\multicolumn{4}{c}{\textbf{Emission Lines}} \\\hline
ha\_br\_*, ha\_nr\_* & -- & group & Broad/narrow \ha \\
hb\_br\_*, hb\_nr\_* & -- & group & Broad/narrow \hb \\
hg\_br\_*, hg\_nr\_* & -- & group & Broad/narrow \hg \\
o\_iii\_4364/5007\_*, o\_i\_6300\_* & -- & group & \oiii $\lambda4364$, \oiii $\lambda5007$ and [O\,\textsc{i}] $\lambda6300$ lines \\
n\_ii\_6549/6585\_* & -- & group & \nii doublet \\
s\_ii\_6718/6732\_* & -- & group & \sii doublet \\
he\_i\_br*, he\_i\_nr* & -- & group & Broad/narrow He\,\textsc{i} $\lambda5875$ line \\
he\_ii\_br*, he\_ii\_nr* & -- & group & Broad/narrow He\,\textsc{ii} $\lambda4686$ line \\\toprule

\multicolumn{4}{c}{\textbf{Ratios and Diagnostics}} \\\hline
hb\_oiii\_ratio       & -- & scalar & log \ratio flux ratio (broad \hb flux only)\\
hb\_oiii\_ratio\_low / upp & -- & error & Uncertainty on the ratio \\
agn\_frac & -- & scalar & AGN power-law to total flux ratio at $\lambda5100$\AA \\
host\_frac & -- & scalar & Host galaxy to total flux ratio at $\lambda5100$\AA \\
fe\_ii\_r\_hb & -- & scalar & Fe\,\textsc{ii} to \hb ratio; Fe\,\textsc{ii} integrated over $\lambda$4340\,\AA\ -- $\lambda$4680\,\AA \\
fe\_ii\_r\_oiii & -- & scalar & Fe\,\textsc{ii} to \oiii ratio; Fe\,\textsc{ii} integrated over $\lambda$4340\,\AA\ -- $\lambda$4680\,\AA\\
fe\_ii\_r\_*\_low / upp & -- & error & Symmetric uncertainty in Fe\,\textsc{ii} ratios; estimated from error spectrum\\\toprule

\multicolumn{4}{c}{\textbf{Continuum Properties}} \\\hline
pl\_s, pl\_s\_low / upp & -- & scalar & AGN power-law spectral index ($\alpha_\nu$) and uncertainty\\
l5100, l5100\_low / upp & log erg s$^{-1}$ & scalar & AGN luminosity at $\lambda$5100\,\AA\ and uncertainty\\
h5100, h5100\_low / upp & log erg s$^{-1}$ & scalar & Host contribution at $\lambda$5100\,\AA\ and uncertainty\\
t5100, t5100\_low / upp & log erg s$^{-1}$ & scalar & Total luminosity at $\lambda$5100\,\AA\ and uncertainty\\\toprule

\multicolumn{4}{c}{\textbf{Black Hole Mass Estimates}} \\\hline
lc\_m       & log $M_\odot$ & scalar & \lcont-based mass estimate \citep{feng_shen_li_2014_l5100_mass} \\
hb\_m       & log $M_\odot$ & scalar & \lhb-based mass estimate \citep{vestergaard_peterson_2006} \\
oiii\_m     & log $M_\odot$ & scalar & \loiii-based mass estimate \citep{greene_hood_2010_oiii_RL}\\
ha\_m     & log $M_\odot$ & scalar & \lha-based mass estimate \citep{cho_woo_2023_halpha_mass}\\
ww\_*       & log $M_\odot$ & scalar & Mass estimates from \citet{amrutha_masscorr_2026} using $R$-$L$ relations in \citet{wang_woo_2024} \\
*\_m\_low / upp & log $M_\odot$ & error & Uncertainty on mass estimates \\\toprule

\multicolumn{4}{c}{\textbf{Data Quality and Fit}} \\\hline
hab\_snr, hbb\_snr, hgb\_snr & -- & scalar & SNR of broad Balmer lines: \ha, \hb, \hg \\
r\_chi, r\_sq & -- & scalar & Reduced $\chi^2$ of the fitted model and $R^2$ statistic \\
median\_snr, blue\_snr, red\_snr & -- & scalar & Median spectrum SNR per \AA\ for full spectrum, blue arm and red arm \\
*\_flag & -- & group & Flags described in Appendix~\ref{appendix_2}\\
bel\_src, cite\_link, cite\_name & -- & group & Reference literature AGN classification\\

\bottomrule
\end{tabular}
\end{table*}

From the spectral decomposition we derive a set of continuum and emission-line measurements for each spectrum. The catalogue includes basic source information (object identifiers, coordinates, redshift, Galactic reddening, and observation date), together with measurements of the continuum and emission-line properties obtained from the BADASS3 fits. Emission-line quantities include luminosities, full width at half-maximum (FWHM; in velocity units), velocity dispersions, and equivalent widths for both broad and narrow components of the Balmer lines, as well as forbidden and high-ionisation transitions including \oiii, \nii, \sii, He\,\textsc{i}, and He\,\textsc{ii}. For each parameter we report the median from the posterior distribution together with the corresponding 16th and 84th percentile uncertainties.

Continuum properties derived from the spectral fits include the AGN power-law slope ($\alpha_\nu = -2-\alpha_\lambda$), the monochromatic continuum luminosity at $5100$~\AA, and the relative contributions of the AGN and host galaxy components at $5100$~\AA. In addition, we provide several diagnostic quantities commonly used in AGN studies, including the \ratio flux ratio, Fe\,\textsc{ii} strength parameters, and signal-to-noise estimates for the principal broad emission lines.

While BADASS3 directly extracts line parameters from the fits, we separately extract the line dispersion of the summed components such that it is calculated within a window defined by $\pm2.5\times$ the median absolute deviation (MAD) around the half-flux wavelength, following \citet{wang_shen_2019}. We also report equivalent widths computed using the AGN power law slope component as the sole continuum, rather than a total continuum that includes the host galaxy and Fe\,\textsc{ii} features.

We provide two sets of single-epoch black hole mass estimates in the catalogue. The first set is based on commonly used virial mass prescriptions, using the relations from \citet{feng_shen_li_2014_l5100_mass}, \citet{vestergaard_peterson_2006}, and \citet{greene_hood_2010_oiii_RL} for \lcont, \lhb, and \loiii, respectively. In all cases, the virial velocity is estimated from the broad \fwhmhb. The second set is based on the radius--luminosity relations from \citet{wang_woo_2024}, adopting the corresponding mass calibrations from \citet{amrutha_masscorr_2026}. In addition, black hole masses derived from the broad \ha luminosity and FWHM are computed using the relation from \citet{cho_woo_2023_halpha_mass}.

Goodness-of-fit metrics from the Bayesian modelling, including reduced $\chi^2$ of the model relative to the data in each wavelength channel, are also reported. The redshifts of the objects are updated using the core component of the \oiii$\lambda5007$\AA\ velocity offset, with typical corrections of $\sim$50~km~s$^{-1}$, capped at $\sim$450~km~s$^{-1}$. The full list of catalogue columns is provided in Table~\ref{tab:extracted_pars}, and the complete catalogue is released in machine-readable format alongside the data products described in Appendix~\ref{appendix_1}.
\section{Quality control and validation}\label{sec:validation}
To quantify the reliability of the released spectra and catalogue products, we perform a series of quality-control and validation tests. These include checks of the emission-line fitting, error-spectrum calibration, source blending, and the assignment of catalogue quality flags. In this section, we go through each of these steps.

\subsection{Broad Balmer emission line fits}

\begin{figure}
    \centering
    \includegraphics[width=1\linewidth]{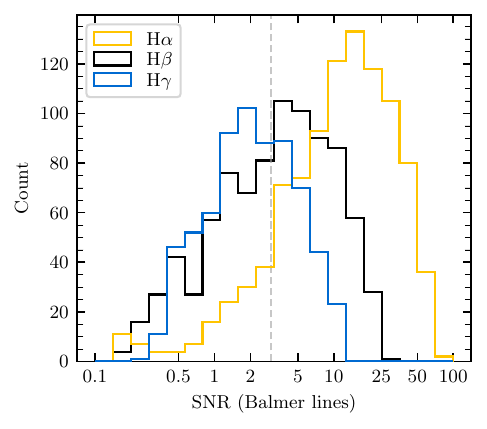}
    \caption{SNR distribution of the fitted broad Balmer lines. The SNRs are computed by dividing the peak of the summed Gaussian components by the mean noise level in the fitted region.}
    \label{fig:snr_br_hist}
\end{figure}

We assess the reliability of the emission-line decomposition using signal-to-noise ratio (SNR) measurements for the broad Balmer components. We define the SNR as the ratio between the peak amplitude of the fitted broad Gaussian components and the mean noise level in the surrounding continuum region. The distributions of the broad-line SNRs are shown in Figure~\ref{fig:snr_br_hist}. Throughout this work, we consider a broad emission line to be undetected when $\mathrm{SNR} < 1$, while robust measurements typically require $\mathrm{SNR} > 3$. In the intermediate regime ($1 < \mathrm{SNR} < 3$), the fits become increasingly uncertain. 

As expected, the higher-order Balmer lines are progressively less well detected. Broad \hb is particularly susceptible to degeneracies with the Fe\,\textsc{ii} pseudo-continuum, the AGN continuum, and broad He\,\textsc{ii} emission. In some cases, these degeneracies cause one of the broad \hb components to become extremely broad and low in amplitude, effectively mimicking a continuum-like feature and driving the fitted FWHM to the imposed upper limit. We identify such cases by selecting fits in which one of the broad \hb components reaches the maximum allowed width. The affected component is then excluded from the broad \hb measurement, which is instead computed using the remaining broad component. To account for the removed flux, the luminosity of the excluded component is added in quadrature to the upper uncertainty of the retained broad \hb luminosity. These objects are flagged using the \texttt{spurious\_hbb\_flag} parameter (see list of flags in Section~\ref{subsec:flags}).

\begin{figure*}
    \centering
    \includegraphics[width=1\linewidth]{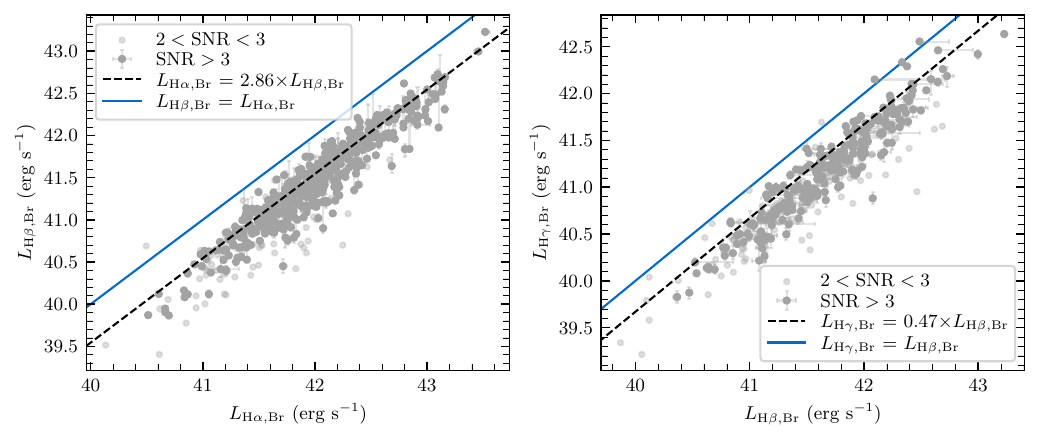}
    \caption{Comparison of the broad component Balmer line luminosity for \ha, \hb and \hg. The solid and dashed lines show the 1:1 relation and Case-B scenario respectively. Different SNR regimes are marked separately. For $\mathrm{SNR}>3$, both lines need to fulfil the condition, but for $2<\mathrm{SNR}<3$, one of the lines may have $\mathrm{SNR}>3$.}
    \label{fig:balmer_lum}
\end{figure*}

\begin{figure*}
    \centering
    \includegraphics[width=1\linewidth]{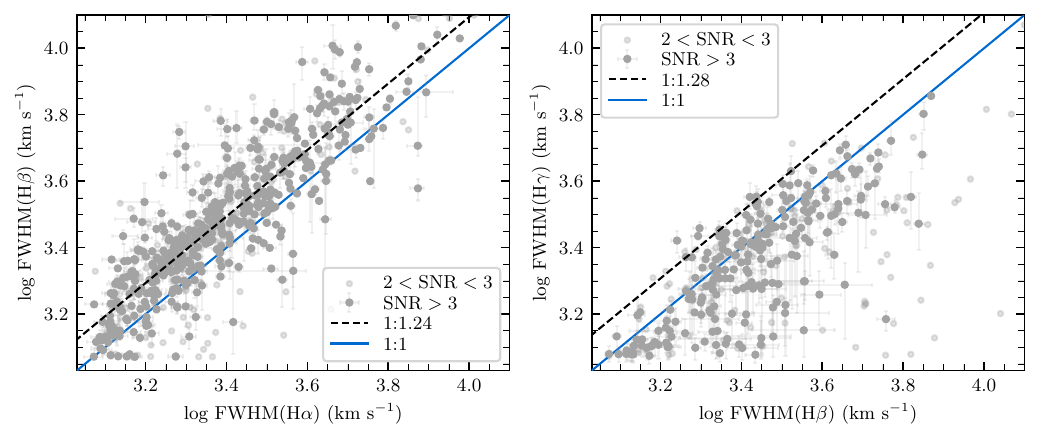}
    \caption{Comparison of the broad component Balmer line FWHM for \ha, \hb and \hg (see Figure~\ref{fig:balmer_lum}  for SNR regimes). The solid line marks the 1:1 relation and the dashed line indicates the expected width relations based on the reverberation-mapping lag ratios of \citet{bentz_walsh_2010}, assuming Keplerian motion ($V^2\propto 1/R$). These predict \fwhmhb = 1.24~$\times$~\fwhmha and \fwhmhg = 1.28~$\times$~\fwhmhb. The observed median ratios are \fwhmhb/\fwhmha~$\sim1.24$, in agreement with the prediction, and \fwhmhg/\fwhmhb~$\sim0.83$, substantially below the expected value.}
    \label{fig:balmer_width}
\end{figure*}

In Figure~\ref{fig:balmer_lum} we compare the luminosities of the broad Balmer lines for objects with $\mathrm{SNR} > 3$ for all lines, and separately mark sources with $2 < \mathrm{SNR} < 3$ for at least one line. The majority of the sample follows the expected trends, with the broad \hb--\ha relation scattering around Case~B recombination \citep[broad \ha/\hb~$=2.86$;][]{dopita_sutherland_2003} and extending to larger ratios as previously observed in AGN BLRs \citep{martin_2017_caseB}. A similar distribution is seen in the broad \hg--\hb relation, with most objects lying below the Case~B expectation (broad \hb/\hg~$=2.12$). 
 
In Figure~\ref{fig:balmer_width} we compare the measured broad-line FWHM values in velocity units. For the objects with reliable measurements, we find the median ratio \fwhmhb/\fwhmha~$\sim1.24$. This is consistent with reverberation-mapping results showing that the broad \hb-emitting region lies closer to the ionising source than the broad \ha-emitting region \citep{bentz_walsh_2010}. Assuming virial motion ($V^2 \propto 1/R$), the measured lag ratio $\tau_{\mathrm{H}\alpha}/\tau_{\mathrm{H}\beta}=1.54$ implies an expected line-width ratio of \fwhmhb/\fwhmha~$\sim1.24$, in excellent agreement with our observations.

A similar comparison between broad \hg and \hb is less robust. We measure a median \fwhmhg/\fwhmhb~$\sim0.83$, whereas the reverberation-mapping lag ratio reported by \citet{bentz_walsh_2010} predicts a value of $\sim1.28$ under the same virial assumption. This discrepancy is likely driven by measurement uncertainties in \hg, an intrinsically weak line that overlaps with host-galaxy absorption features and is blended with the \oiii$\lambda4343$ emission line. Consequently, separating the narrow and broad \hg components is considerably more challenging than for \ha or \hb. This tendency for the broad \hg component to be fitted with a narrower width than expected may contribute, at a sub-dominant level, to the slightly lower \hg/\hb luminosity ratios observed in Figure~\ref{fig:balmer_lum}. However, the deviation from the expected Case~B values is small, and the \hg/\hb ratios remain broadly consistent within the overall scatter. Relative to \hb/\ha, the effect is marginal, suggesting that any bias in the broad \hg flux is likely minor.

\subsection{Error spectra}
The propagated error spectra are not always reliable due to the presence of bad pixels in the data cubes. In particular, individual spaxels in regions dominated by sky emission can contain pixels with artificially large uncertainties, occasionally reaching values of order $\sim10^8$ times the median error. These extreme values are typically associated with flags created during \texttt{PyWiFeS} reduction due to detector artefacts. When constructing the median sky spectrum and propagating uncertainties through the sky-subtraction process, such outliers can disproportionately inflate the resulting error spectrum, even when the underlying flux measurements are otherwise well behaved.
To mitigate this effect, we impose a conservative cap on the error spectrum. Specifically, any pixel in the error spectrum with a value exceeding 20 times the median of the error spectrum is reset to the median error value. This approach preserves the global noise characteristics of the data while preventing a small number of pathological pixels from dominating the overall uncertainty estimates. Hence, we quote medians rather than mean errors and SNRs for spectra as a whole. In turn, this stabilises subsequent spectral decomposition, where inflated errors can otherwise bias the weighting of spectral regions and degrade the quality of the fits.

\subsection{Source blending}
We cross-match each source with the nearest objects in the SMSS DR4 and Gaia DR3 catalogues to identify potential counterparts and blended sources. For each source, we report the closest match within a search radius of 15 arcsec, along with its angular separation. Given the 6.7 arcsec diameter extraction aperture used to construct the spectra, sources separated by less than 3.35 arcsec are expected to be centred within the aperture, and along with the PSF, a larger separation is required to ensure blending does not affect the extracted spectrum. In such cases, flux from neighbouring objects contributes to the observed spectrum. Given a seeing range from 1 arcsec to 3 arcsec in the observations, we expect noticeable contamination of the spectra from point sources separated as far away as 6 arcsec and potentially from extended sources that are even further away. 

This blending predominantly affects the continuum emission, as emission lines, particularly broad lines, are less susceptible to contamination unless the secondary source is comparably bright. However, when the contaminating source contributes significantly to the continuum, the model fails to explain the observed spectrum, potentially leading to unreliable continuum parameters or decompositions. Neighbouring galaxies at nearly identical redshifts primarily alter the mix of stellar populations contributing to the host-galaxy spectrum. In contrast, projected neighbours at substantially different redshifts can introduce a second set of spectral features, potentially complicating line identification and measurements. Contamination from foreground stars is generally easier to identify, as stellar absorption features at $z=0$ typically stand out against the AGN spectrum.

\subsection{Catalogue flags}\label{subsec:flags}
To assist in assessing the reliability of the derived spectral measurements, we provide a set of diagnostic quality flags, described in Appendix~\ref{appendix_2}. These flags capture common fitting issues and edge cases in a uniform and reproducible manner, enabling users to construct clean subsamples without inspecting individual spectra in detail. They are derived directly from the catalogued spectral properties and are intended as a general guide to measurement reliability. Although flag-based selections can be tailored to specific scientific applications, we recommend restricting samples to sources with individual flag values of 0 or 1 for general use, balancing measurement reliability against sample size.

\subsection{Seyfert classification}\label{subsec:sytype}

\begin{figure*}
    \centering
    \includegraphics[width=0.9\linewidth]{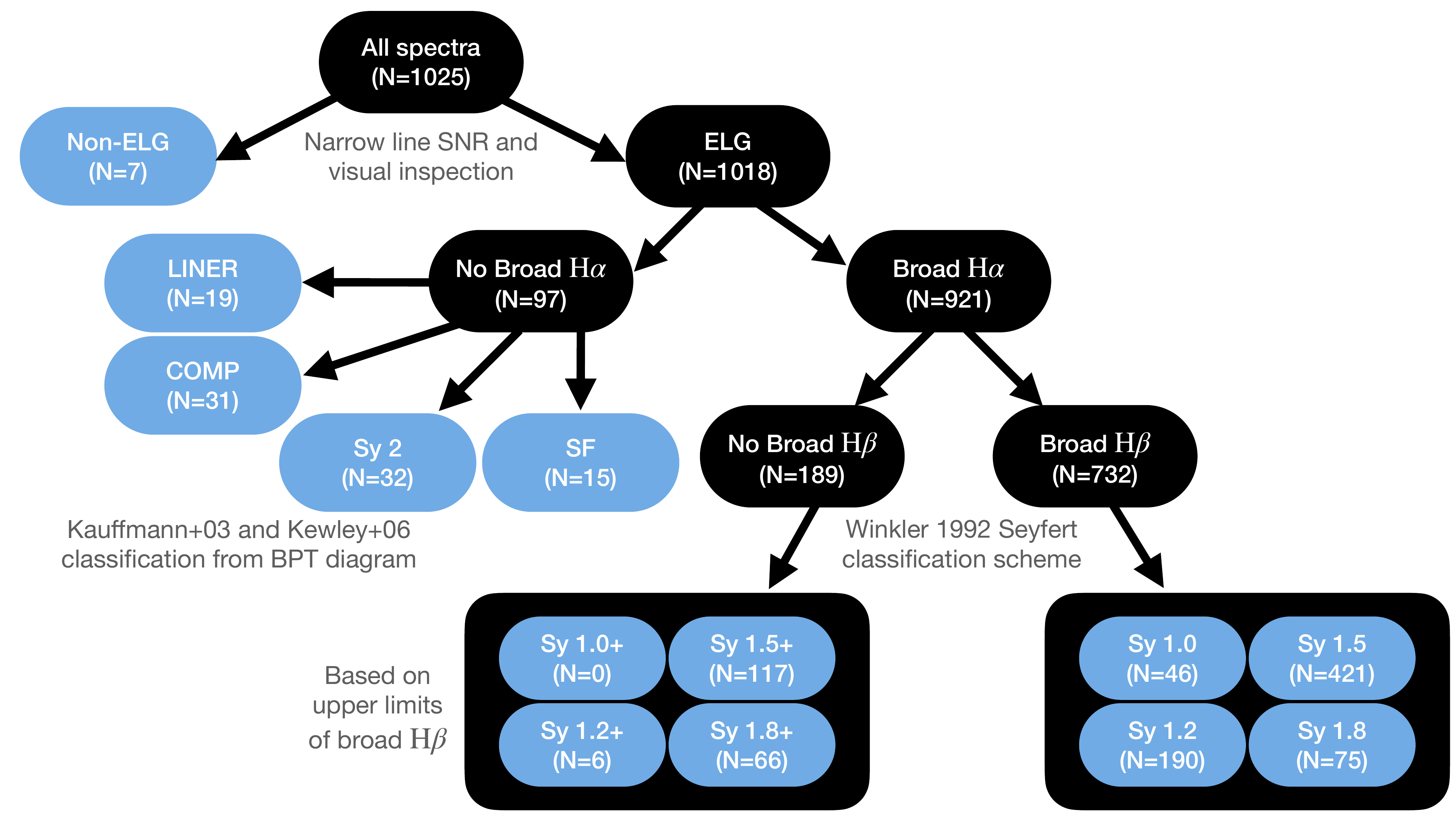}
    \caption{Spectrum classification decision tree. The full procedure is described in Section~\ref{subsec:sytype}. The end nodes of the tree denote classification labels presented in the \texttt{type} column of the catalogue. The 30 spectra with poor data quality are flagged as \texttt{Bad} and are not assigned a spectral type. For objects with multiple observations, all epochs not flagged as \texttt{Bad} are included.}
    \label{fig:flowchart}
\end{figure*}

The catalogue includes a \texttt{type} column that provides a spectroscopic classification for each object. 30 objects with bad spectrum quality are labelled \texttt{Bad} and are not assigned a classification. Classification follows a decision tree (see Figure~\ref{fig:flowchart}) based on the detection of narrow and broad emission lines. We first identify emission-line galaxies (ELGs) using the narrow \ha, \oiii, \nii, and [S\,\textsc{ii}] lines. Objects with no significant detection of any of these features are classified as \texttt{Non-ELG}, following visual inspection of cases where the individual narrow-line SNRs are below unity. For ELGs without a significant broad \ha component, we apply standard Baldwin–Phillips–Terlevich \citep[BPT;][]{baldwin_phillips_BPT_1981} diagnostics to classify Seyfert~2 (\texttt{Sy 2}), star-forming (\texttt{SF}), composite (\texttt{COMP}), and low-ionization narrow emission-line regions (\texttt{LINER}) following the demarcations of \citet{kauffmann_heckman_2003} and \citet{kewley_groves_2006}.

ELGs with a broad \ha detection (\texttt{hab\_snr}~$\geq 1$) are classified as broad-line AGN. These objects are further subdivided using the broad \hb emission. Where a reliable broad \hb measurement is available (\texttt{hbb\_snr}~$\geq 1$), we assign Seyfert subtypes (\texttt{Sy 1.0}, \texttt{Sy 1.2}, \texttt{Sy 1.5}, and \texttt{Sy 1.8}) following the line-ratio scheme of \citet{winkler_1992}, based on the \ratio ratio. When broad \hb is not reliably detected (\texttt{hbb\_snr}~$<1$ with \texttt{hab\_snr}~$>1$), we do not assume its absence. Instead, we compute an upper limit on the broad \hb flux by adopting a Gaussian profile with peak amplitude set to the local noise level and a width tied to the broad \ha component via the expected virial scaling (\fwhmhb/\fwhmha~$\sim1.24$). These objects are assigned extended Seyfert classes (\texttt{Sy 1.0+} to \texttt{Sy 1.8+}) that replace the traditional Seyfert~1.9 category but explicitly reflect the possibility that broad \hb may be present below the detection threshold rather than physically absent \citep{bianchi_chiaberge_2022_ngc3147,barquingonzalez_mateos_2024}. The upper limits of the broad \hb luminosity and \ratio ratio are provided in upper error columns of the catalogue (\texttt{hb\_br\_lum\_upp} and \texttt{hb\_oiii\_ratio\_upp} respectively). We note that the quality flags from Section~\ref{subsec:flags} that require \texttt{hbb\_snr}~$>1$ will always have a value of 2 by design for these objects.

\section{Sample properties}\label{sec:sample_props}

In this section, we present an overview of the key properties of the sample and assess the physical consistency of the measurements. We examine emission-line diagnostics using the Baldwin–Phillips–Terlevich diagram, investigate dust attenuation through the Balmer decrement, and explore continuum scaling relations relevant to AGN luminosity. Together, these properties provide a global view of the ionisation conditions, line-emitting regions, and continuum properties of the sample, while also serving as a validation of the spectral decomposition. We conclude with a brief discussion of a few individual objects in this sample.

\subsection{Ionisation diagnostics}

\begin{figure*}
    \centering
    \includegraphics[width=1\linewidth]{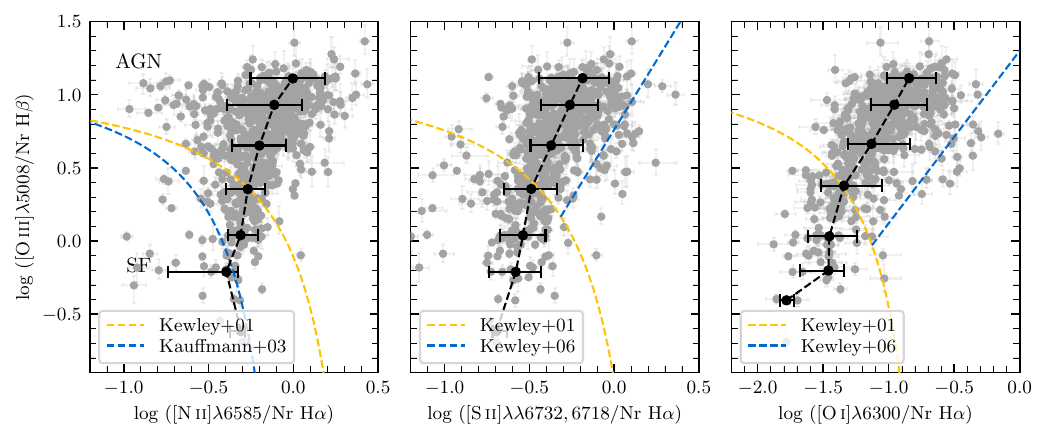}
    \caption{BPT diagram showing narrow line ratios log \oiii$\lambda5008$/\hb versus log \nii$\lambda6585$/\ha, log [S\,\textsc{ii}]$\lambda\lambda6732,6718$/\ha, and log \oiii$\lambda6300$/\ha. The dashed curves in the \nii panel indicate the demarcation lines from \citet{kewley_heisler_2001}, \citet{kewley_groves_2006} and \citet{kauffmann_heckman_2003} separating star forming and AGN regions. The regions between the curves are typically classified as composite galaxies. The remaining panels only include the \citet{kewley_heisler_2001} line. In each panel, a running median relation is shown, with error bars indicating the 16th–84th percentile range. For visual clarity, only 70 per cent of the sample with the least measurement errors are presented here. The remaining sample follows the same distribution but have larger errors. The \citet{kauffmann_heckman_2003} line is used to classify Seyfert 2 and star forming galaxies, and the \citet{kewley_groves_2006} line is used to identify low-ionization narrow emission-line regions, as described in Section~\ref{subsec:sytype}.}
    \label{fig:bpt}
\end{figure*}

We present the standard BPT diagnostic diagrams in Figure~\ref{fig:bpt}, showing the narrow-line ratios \oiii$\lambda5008$/\hb as a function of \nii$\lambda6585$/\ha, [S\,\textsc{ii}]$\lambda\lambda6732,6718$/\ha, and \oiii$\lambda6300$/\ha for the sample. We also indicate the theoretical and empirical boundaries separating AGN, composite, and star-forming regions from \citet{kewley_heisler_2001} and \citet{kauffmann_heckman_2003}. The majority of sources lie within the AGN-dominated region, as expected from the selection of the parent sample. A smaller fraction extends towards the composite region, which may reflect residual host-galaxy contamination or intrinsically weak AGN activity. 

The overall distribution is consistent with previous low-redshift AGN samples, indicating that the emission-line measurements and spectral decomposition recover meaningful line ratios across the sample. In particular, the separation of broad and narrow components is verified with the BPT diagram.

\subsection{Balmer decrement}

\begin{figure*}
    \centering
    \includegraphics[width=1\linewidth]{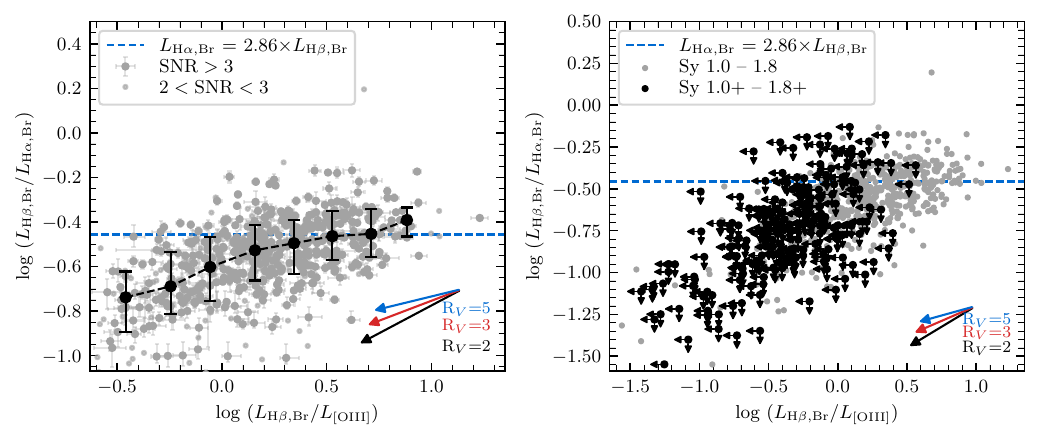}
    \caption{Balmer decrement \lhb/\lha as a function of the \ratio ratio. The running median is shown, with error bars indicating the 16th–84th percentile range. The horizontal line marks the Case~B recombination value. \textbf{Left:} Sample limited to objects with broad \ha and \hb SNR~$>2$. The SNR regimes are marked separately, following from Figure~\ref{fig:balmer_lum}. \textbf{Right:} The full sample, separating between the Seyfert types 1.0 to 1.8 (\hb SNR~$>1$) and the sample limited to Seyfert types 1.0+ to 1.8+ with upper limits for the \hb detection, traditionally the Seyfert type 1.9s  (\hb SNR~$<1$). The annotated arrows mark reddening vectors from \citet{calzetti_armus_2000} for $A_V=1$~mag.}
    \label{fig:decrement}
\end{figure*}

In Figure~\ref{fig:decrement} we compare the broad Balmer decrement, \lhb/\lha, against the \ratio flux ratio. The Balmer decrement is sensitive to dust attenuation along the line of sight, and also BLR optical depth and ionisation conditions \citep{martin_2017_caseB}, while the \ratio ratio traces the relative strength of BLR and narrow-line region (NLR) emission. The latter is sensitive to viewing angle, anisotropic emission, variability on timescales shorter than \oiii reverberation, and dust obscuration. The distribution shows a broad correlation with significant scatter. Objects with lower broad Balmer decrement ratios (indicative of increased reddening) tend to exhibit lower \ratio values, consistent with preferential attenuation of the BLR emission relative to the more extended NLR. In Figure~\ref{fig:decrement}, reddening vectors are shown for $A_V=1$~mag for $R_V=$~\{2, 3, 5\} under the assumption that the dust reddens only the broad Balmer lines and leaves the \oiii emission from the extended narrow-line region unaffected. The overall distribution is broadly consistent with an extinction vector corresponding to $R_V\sim2$, although its shape may reflect a combination of dust reddening and variations in AGN activity, rather than a single reddening sequence. However, the observed scatter likely reflects a combination of dust geometry, intrinsic variability of the broad-line emission, and differences in the relative contributions of the BLR and NLR. Disentangling these effects requires multi-epoch or multi-wavelength observations; the catalogue presented here provides a starting point for identifying suitable targets for such studies.

\subsection{Continuum scaling relations}\label{sec:cont_sci}

\begin{figure}
    \centering
    \includegraphics[width=1\linewidth]{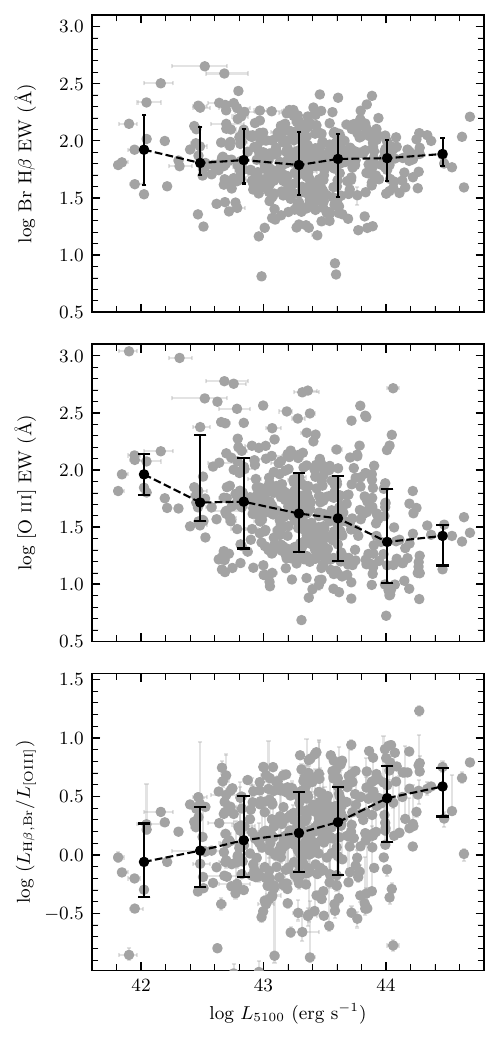}
    \caption{Equivalent width of broad \hb and narrow \oiii, along with the \ratio luminosity ratio as a function of AGN continuum luminosity, \lcont. Running medians are shown, with error bars indicating the 16th–84th percentile range. A Baldwin effect is evident for narrow \oiii, while the trend is absent for broad \hb.}
    \label{fig:baldwin}
\end{figure}

We investigate the relation between emission-line equivalent width and continuum luminosity in Figure~\ref{fig:baldwin}. The Baldwin effect \citep{baldwin_1977} predicts an anti-correlation between line equivalent width and continuum luminosity, such that more luminous AGN exhibit weaker emission lines. However, the strength of this relation is expected to differ between emission lines due to their distinct physical origins. The \oiii emission arises in the spatially extended narrow-line region and traces the time-averaged ionising luminosity of the AGN, whereas the broad \hb emission originates in the compact broad-line region and responds to continuum variations on much shorter timescales. As a result, single-epoch measurements of broad \hb are more strongly affected by intrinsic variability, and are therefore expected to exhibit a weaker Baldwin relation compared to \oiii. Consistent with this expectation, we observe a clear anti-correlation for the \oiii emission, in agreement with previous studies of AGN narrow lines \citep{zhang_wang_2013}, while the broad \hb emission shows only a weak dependence on continuum luminosity. This discrepancy in the strength of the Baldwin effect between broad \hb and narrow \oiii has important implications for single-epoch black hole mass estimates, as discussed in \cite{amrutha_masscorr_2026}.

\begin{figure*}
    \centering
    \includegraphics[width=0.98\linewidth]{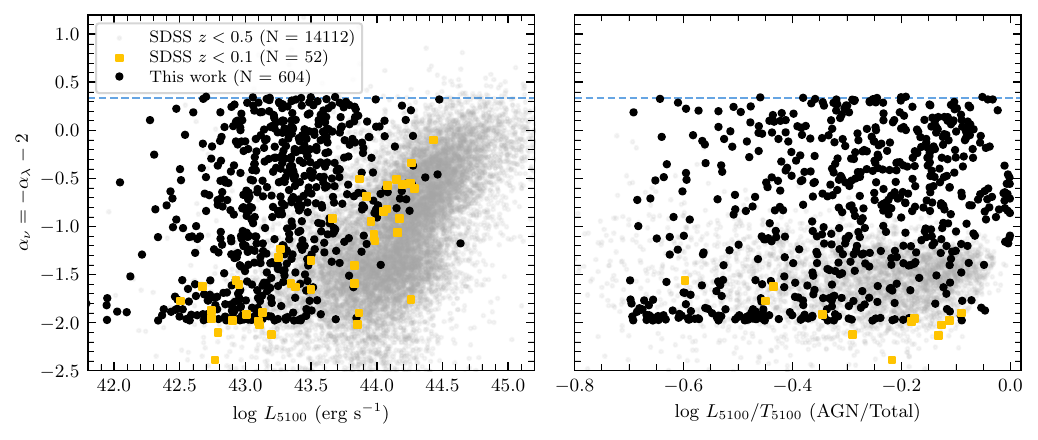}
    \caption{AGN power-law spectral index (in frequency space) as a function of \textbf{(left)} continuum luminosity, \lcont, and \textbf{(right)} the AGN fraction of the total continuum. The horizontal line marks the theoretical expectation for a standard thin accretion disc ($\alpha_\nu = 1/3$). A comparison with the SDSS quasar sample from \citet{rakshit_stalin_2020} is also presented. Objects with only AGN contribution and no detected host components from the SDSS data set are omitted from the right panel, as the SDSS data provide null values for the host fractions.}
    \label{fig:pls}
\end{figure*}

In Figure~\ref{fig:pls}, we present the fitted AGN power-law spectral index as a function of the AGN luminosity, \lcont, and the AGN fraction of the total continuum. To focus on reliable measurements, we exclude values that lie outside the plausible range and those clustered at the imposed boundary, restricting the sample to $-1.95 < \alpha_\nu < 0.35$ and a minimum AGN fraction of 0.2. This selection retains approximately 60 per cent of the sample. We compare our results to the SDSS quasar sample analysed by \citet{rakshit_stalin_2020}, who also perform host–AGN spectral decomposition. We limit the SDSS sample to $z<0.5$. Although the SDSS $z<0.1$ sample overlaps with the higher luminosity end of our sample, a correlation is observed between spectral slope and luminosity in the SDSS sample, whereas only a weak trend is present in our sample. Especially, our atlas contains many lower-luminosity objects with clearly blue continua.

The discrepancy can be understood in terms of the treatment of the host galaxy component. Typical host galaxy fitting procedures do not fully account for the additional reddening of host galaxies due to dust, effectively limiting the range of allowed host spectral shapes. This introduces a degeneracy in which intrinsically red host galaxies are instead absorbed into the AGN component, leading to artificially redder inferred AGN slopes at lower luminosities. When we restrict our own modelling to exclude this additional degree of freedom, we recover a similar slope–luminosity correlation observed in the SDSS sample, confirming that the effect is driven by modelling assumptions rather than intrinsic AGN physics. In the extreme case where no host component is included, the typically red host contribution naturally biases lower-luminosity AGN towards redder inferred continua.

This interpretation is further supported by the right panel of Figure~\ref{fig:pls}. For sources with significant host contributions, the SDSS sample occupies only the red half of the plausible spectral index range, whereas our decomposition allowing for dust-reddened hosts, recovers the full range of spectral slopes. Conversely, the SDSS sample does not identify host components for many objects with bluer spectral indices. We note, however, that differences in extraction aperture between this work (6.7 arcsec) and the SDSS analysis (3 arcsec) may also contribute to these trends. These results highlight the challenges in reliably separating AGN and host galaxy continua. Care must therefore be taken when interpreting spectral slope measurements from single-epoch optical spectra, particularly in samples where host contamination is significant.

\subsection{Notes on special objects}

\subsubsection{Non-emission-line galaxies}

We identify seven objects with spectra lacking significant emission-line signatures: 
\textit{LEDA 158271, LEDA 156650, LEDA 135030, PKS 2156$-$203, LEDA 131166, PKS J2231-0824}, and \textit{Fairall 524}. All seven sources are classified as type 1 AGNs in the Milliquas catalogue, with spectroscopic references traced to \citet{paturel_petit_2003_pgc}. However, the WiFeS spectra presented here do not exhibit the broad or narrow emission lines typically associated with AGN activity. Weak narrow \nii and \oiii emission is detected in \textit{PKS J2231$-$0824}, but the source remains dominated by stellar continuum features. For the purposes of this atlas, these objects are classified as non-emission-line galaxies (\texttt{Non-ELG}).

\subsubsection{Non-Gaussian line profiles}
Four objects exhibit clear double-peaked broad-line profiles: \textit{Ark 120}, \textit{6dFGS J100202.93$-$302117.0}, \textit{HE 0403-3719}, and \textit{ESO 116$-$G010}. These sources were modelled with three broad Gaussian components. The fitting procedure was initialised such that one component was centred on the systemic velocity, while the remaining two components were offset by $\pm1000~\mathrm{km,s^{-1}}$, to allow convergence on the observed double-peaked profiles. A further 25 objects display possible double-peaked features, although these are less pronounced and were adequately modelled using the standard two-component broad-line prescription.

Two sources, \textit{HE 1029-1401} and \textit{LEDA 225607}, were fit with Lorentzian components instead of Gaussian components, after the fitting routine consistently failed with Gaussian components. Two sources, \textit{Mrk 609} and \textit{6dFGS gJ053744.8-570956}, were fit with three broad Gaussian components, and two additional sources, \textit{HE 2302-0857} and \textit{PBC J0917.2-6454}, were fit with three Lorentzian components. However, these objects do not exhibit clear double-peaked profiles. The additional component was introduced because the fitting routine failed to converge using the standard two-component model.

\subsubsection{Miscellaneous}
The 6dFGS source \textit{gJ231809.4$-$112546} is flagged as \texttt{Bad}, not because of poor spectral quality, but because its original 6dFGS classification was incorrect: the broad \hb and \hg lines were misidentified as \ha and He~\textsc{i}, respectively. The source has $z\sim0.4$, but is retained in the atlas to document this misclassification for future reference. For the 6dFGS source \textit{gJ053048.2$-$044044}, we adopt the Milky way extinction, E$(B-V)$, value from \citet{planck_gnilc_dust_2016} instead of the \citet{dust_map_SandF_2011} values used for the rest of the sample. This source is the only one for which the two maps differ by more than 0.25 mag; for all other objects, the absolute differences are below 0.05 mag and can be accommodated by the adopted host-galaxy extinction tolerance.

\subsubsection{Changing-look AGN}

The atlas includes changing-look AGN (CLAGN) identified by \citet{wolf_golding_2020}, \citet{hon_wolf_2022_skymapper_colors}, and \citet{amrutha_wolf_2024}. These represent all systematically identified CLAGN currently known within the selection limits of the atlas.

Most CLAGN are recognised through the appearance or disappearance of broad \hb emission, corresponding to transitions between Seyfert subtypes. Because our sample includes AGN exhibiting broad \ha emission, it is expected to contain the majority of CLAGN transitioning from type 1.9 to type 1.0, 1.2, 1.5, or 1.8. In contrast, transitions from type 2 AGN into broad-line states appear to be intrinsically rare; only one such object was identified among more than 1000 type 2 AGN examined by \citet{amrutha_wolf_2024}, particularly because most of these type 2 AGNs have their broad line regions obscured by the dusty torus rather than being ``true'' type 2s with an intrinsic absence of broad emission. Furthermore, as discussed in Section \ref{subsec:sytype}, the transitions to and from type 1.9/2 states may just be variability across the detection threshold of the broad Balmer lines that is dependent on the instrument resolution.

Objects classified as Seyfert 2 or star-forming galaxies in the present catalogue were previously classified as type 1 AGN in either 6dFGS or Milliquas. In some cases these sources may represent unrecognised CLAGN that have transitioned into a low-accretion state since the earlier observations. Alternatively, the higher spectral resolution and signal-to-noise ratio of the WiFeS observations may allow features previously interpreted as broad emission to be more accurately decomposed into narrow-line components. We therefore retain these objects in the atlas, even when no significant broad-line emission is detected in the WiFeS epoch, as they may provide valuable targets for future variability and CLAGN studies.

\subsection{Example science cases}\label{sec:science}

The primary goal of this atlas is to provide a contemporary spectroscopic reference for broad line AGN in the Southern sky at $z < 0.1$, enabling both variability studies and population-level analyses.
The first motivation for constructing the atlas was quantifying long-term variability of the broad H$\beta$ emission line over a $\sim$20-year baseline relative to earlier 6dFGS observations. This is used in both the \ratio ratio underpinning Seyfert sub-types \citep{winkler_1992} and in estimating black-hole masses. The analysis by \citet{amrutha_masscorr_2026} found evidence for long-term mean-reverting behaviour in the \ratio ratio, where individual objects typically change by up to a factor of $\sim2$, while the sample as a whole remains bounded. These changes are accompanied by corresponding variations in luminosity and inferred single-epoch black hole masses, while broad-line widths remain largely unchanged. Such behaviour is not consistent with simple broad-line region ``breathing'' expectations and suggests that the effective BLR size does not respond strongly to luminosity variations on decadal timescales. Consequently, virial mass estimates based on instantaneous luminosities may be systematically biased by variability, whereas estimates based on longer-timescale luminosity tracers, such as \oiii $\lambda5007$, appear to be more stable.

A photometric variability study of this sample has been carried out by \citet{tan_wolf_2026}, who analysed the ensemble variability through structure functions derived from NASA/ATLAS light curves. Contrary to earlier work over shorter time spans, they found that variability amplitudes rose over the time span probed instead of converging to a decorrelation limit as predicted. Splitting the variability diagnostics by \ratio ratio also suggested that the variety of Seyfert subtypes cannot be explained by an obscuration sequence. 

Disentangling the contributions of accretion variability and variety in obscuration to the observed diversity of AGN subtypes requires coordinated multi-wavelength observations. Roughly two-thirds of the AGN in this atlas have X-ray data from eROSITA, providing an opportunity to investigate the relationship between optical spectral changes and X-ray absorption, including through column densities inferred from X-ray spectral fitting. The atlas also provides a reference against which future transient discoveries can be assessed: spectra obtained following photometric identification of a transient can be compared with the archival observations to identify and characterise spectral changes relative to the underlying AGN. By combining these archival optical spectra with ongoing time-domain surveys and multi-wavelength observations, the atlas will support investigations into the physical mechanisms driving AGN variability, changing-look phenomena, and the diversity of observed AGN subtypes.

\section{Summary}\label{sec:summary}

We present a spectroscopic atlas and catalogue of 1001 broad-line AGNs in the Southern sky at $z < 0.1$, declination $\delta < 0\degree$, and Galactic latitude $|b| > 10\degree$, observed with the WiFeS integral field spectrograph on the ANU 2.3~m telescope. The sample is drawn from 6dFGS \citep{hon_webster_2025}, S7 \citep{thomas_dopita_2017}, AllBRICQS \citep[][Onken et al., in prep]{onken_wolf_AllBRICQS_2023}, and Milliquas \citep{milliquas_2023}, and is designed to be as complete as possible for luminous type 1 AGNs in this parameter space ($K<13$~mag), with residual incompleteness driven by host galaxy dilution at $z\lesssim0.03$. The final dataset comprises of 1055 spectra of the 1001 AGNs, with a median signal-to-noise ratio of 13 and 23 per \AA\ in the blue and red arms respectively. For each source, we provide AGN spectra, AGN--host decomposition, and a uniform catalogue of emission-line properties, kinematics, and continuum luminosities. A small number of objects remain pending observation due to insufficient-quality existing spectra or unfavourable observing constraints, such as proximity to the Sun. Observations of these sources are currently underway and the resulting spectra and measurements will be incorporated into future updates of the online resources for the atlas.

This atlas is designed as a reference dataset for studies of AGN population demographics and variability in the local Universe. The sample overlaps extensively with the footprint of already ongoing large photometric time-domain surveys such as Asteroid Terrestrial-impact Last Alert System and Zwicky Transient Facility. The Legacy Survey of Space and Time (LSST) carried out at the Vera C. Rubin Observatory  will monitor these AGNs photometrically over the coming decade. Beyond characterising stochastic variability, these surveys will enable the identification of transient phenomena such as TDEs and changing-look behaviour in these sources with the spectra presented here providing a reference, allowing comparisons between spectroscopic states before and after photometric triggers. Complementary spectroscopic epochs will be provided by SDSS-V and the 4MOST Hemisphere Survey, both of which are expected to observe a substantial fraction of the sample. Together, these datasets will support systematic studies of long-term spectroscopic variability and AGN demographics across the full Southern broad-line population.

\section*{Acknowledgements}
The authors thank the referee, Hartmut Winkler, for suggesting significant improvements to the manuscript.
NA was supported by Australian Government Research Training Program Scholarship.
NA thanks Ashley Hai Tung Tan and Grace Blomfield for their contributions to improving the manuscript and spectrum decompositions.
This paper is based on data acquired at the ANU 2.3-metre telescope. The automation of the telescope was made possible through an initial grant provided by the Centre of Gravitational Astrophysics and the Research School of Astronomy and Astrophysics at the Australian National University and through a grant provided by the Australian Research Council through LE230100063. The Lens proposal system is maintained by the AAO Research Data \& Software team as part of the Data Central Science Platform. We acknowledge the traditional custodians of the land on which the telescope stands, the Gamilaraay people, and pay our respects to elders past and present. We thank the WiFeS observers Katie Auchettl, Patrick Tisserand and Harrison Abbot for efforts in acquiring the spectra for this paper.
The national facility capability for SkyMapper has been funded through ARC LIEF grant LE130100104 from the Australian Research Council, awarded to the University of Sydney, the Australian National University, Swinburne University of Technology, the University of Queensland, the University of Western Australia, the University of Melbourne, Curtin University of Technology, Monash University, and the Australian Astronomical Observatory. SkyMapper is owned and operated by The Australian National University’s Research School of Astronomy and Astrophysics. The survey data were processed and provided by the SkyMapper Team at ANU. The SkyMapper node of the All-Sky Virtual Observatory (ASVO) is hosted at the National Computational Infrastructure (NCI). Development and support of the SkyMapper node of the ASVO has been funded in part by Astronomy Australia Limited (AAL) and the Australian Government through the Commonwealth’s Education Investment Fund (EIF) and National Collaborative Research Infrastructure Strategy (NCRIS), particularly the National eResearch Collaboration Tools and Resources (NeCTAR) and the Australian National Data Service Projects (ANDS).

\section*{Data Availability}
The final data products as decribed in Appendix \ref{appendix_1} are publicly available at \href{https://www.mso.anu.edu.au/stromlo_agn/atlas/}{https://www.mso.anu.edu.au/stromlo\_agn/atlas/}. SMSS data used in this paper are available at the SkyMapper node of the All-Sky Virtual Observatory (ASVO) at \href{https://skymapper.anu.edu.au}{https://skymapper.anu.edu.au}. The 6dFGS data are available at \href{http://www-wfau.roe.ac.uk/6dFGS/}{http://www-wfau.roe.ac.uk/6dFGS/} and the Final Data Release is available for public access.

\bibliographystyle{mnras}
\input{output_arxiv}

\appendix
\section{Data Structure}\label{appendix_1}
\begin{figure*}
    \centering
    \includegraphics[width=0.9\linewidth]{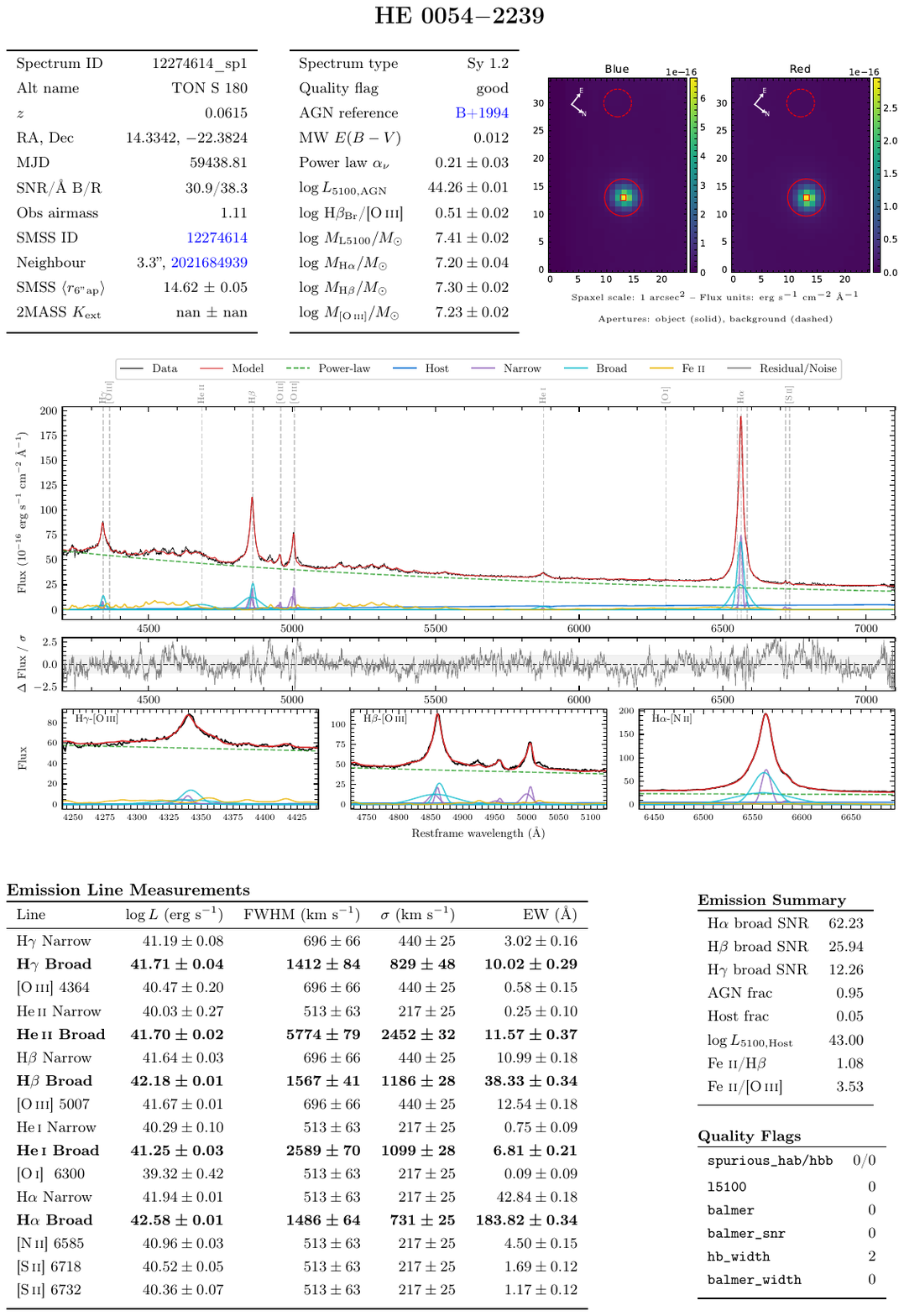}
    \caption{Example summary page for a spectrum.}
    \label{fig:pdf_page}
\end{figure*}

The data can be accessed via the provided web link \href{https://www.mso.anu.edu.au/stromlo_agn/atlas/}{https://www.mso.anu.edu.au/stromlo\_agn/atlas/}. The data is presented as follows:
\begin{itemize}
    \item \texttt{catalogue.csv} (file): The main catalogue of derived parameters for all sources. Column definitions are provided in Table~\ref{tab:extracted_pars}.
    \item \texttt{raw\_spectra/} (directory): FITS files containing the extracted 1D spectra.
    \item \texttt{decomp\_spectra/} (directory): Spectral decomposition output from \textsc{BADASS3}, provided in tabular format. Fluxes are reported in units of $10^{-16}$ erg s$^{-1}$ cm$^{-2}$ \AA$^{-1}$.
    \item \texttt{pdf\_pages/} (directory): Per-object summary pages in PDF format, including the spatial extraction image, the decomposed spectrum, and derived parameters. An example is shown in Figure~\ref{fig:pdf_page}.
    \item \texttt{object\_pages/} (directory): HTML summary pages for each object, containing the same information as \texttt{pdf\_pages} together with an interactive spectrum viewer.
    \item \texttt{merged\_summaries.pdf} (file): A compiled PDF containing all individual pages from the \texttt{pdf\_pages/} directory ($\sim0.5$ GB).
\end{itemize}

Files are named using the corresponding \texttt{spectrum\_id} in \texttt{catalogue.csv} in the directories. The base web page includes a table for quick access of the individual objects.

\section{Quality flag definitions}\label{appendix_2}

Here, we list all quality flags used in the catalogue:
\begin{itemize}
\item \texttt{l5100\_flag}: quantifies the reliability of the AGN continuum luminosity at 5100\,\AA. This flag combines constraints on the spectral slope, luminosity, and fractional AGN contribution. A value of 0 corresponds to high-confidence measurements, requiring a physically plausible spectral index \citep[$-2 < \alpha_\nu \leq 0.5$;][]{xie_shao_2016,rakshit_stalin_2020}, sufficient luminosity ($\log L_{5100} > 41.5$), and a dominant AGN contribution ($\mathrm{agn\_frac} \geq 0.5$). A value of 1 indicates moderate reliability, where the spectral shape and luminosity appear acceptable but the AGN contribution is lower ($0.2 < \mathrm{agn\_frac} < 0.5$), increasing the potential impact of host-galaxy contamination. A value of 2 flags the remaining cases, typically associated with unphysical slopes, low luminosities, or strong degeneracies in the continuum decomposition. The flag values are roughly split into thirds for this criterion.
\item \texttt{balmer\_flag}: assesses the physical consistency of the broad Balmer line luminosities. Under standard conditions, the broad \hb luminosity is expected to be lower than that of \ha, although we note that we are not including the Case-B cut of \ha $> 2.86\times$ \hb. A value of 0 corresponds to the physically consistent regime with a robust detection ($\mathrm{SNR}_{\mathrm{H}\beta} > 3$). A value of 1 indicates marginal detections, where $1 < \mathrm{SNR}_{\mathrm{H}\beta} \leq 3$, and the inferred luminosity ratio is therefore less certain. A value of 2 flags cases that are either physically inconsistent ($L_{\mathrm{H}\beta} \geq L_{\mathrm{H}\alpha}$) or have very low SNR, both of which suggest unreliable line measurements. The flag values \{0, 1, 2\} are roughly split into \{0.50, 0.25, 0.25\} for this criterion.
\item \texttt{balmer\_snr\_flag}: provides a summary measure of the overall quality of the broad Balmer-line detections. A value of 0 indicates robust detections in all three lines (\ha, \hb, and \hg), with $\mathrm{SNR} > 3$ in each case. A value of 1 corresponds to intermediate-SNR detections, requiring $\mathrm{SNR}_{\mathrm{H}\alpha} > 3$, $\mathrm{SNR}_{\mathrm{H}\beta} > 2$, and $\mathrm{SNR}_{\mathrm{H}\gamma} > 1$, thereby allowing for progressively lower SNR in the weaker lines. A value of 2 flags low-SNR detections where one or more of the lines fall below these thresholds. The flag values \{0, 1, 2\} are roughly split into \{0.25, 0.25, 0.50\} for this criterion, although we note that this flag depicts the broad Balmer emission strength more than the quality of the spectrum.
\item \texttt{hb\_width\_flag}: evaluates the internal consistency of the broad \hb line-width measurements by comparing the logarithmic FWHM and velocity dispersion ($\sigma_\mathrm{line}$). For a well-measured line profile, these quantities are expected to follow a relatively narrow relation. A value of 0 corresponds to $0.3 < \log \mathrm{FWHM}_{\mathrm{H}\beta} - \log \sigma_{\mathrm{H}\beta} \leq 0.5$, consistent with typical broad-line profiles. A value of 1 allows for a wider range ($0.15 < \log \mathrm{FWHM}_{\mathrm{H}\beta} - \log \sigma_{\mathrm{H}\beta} \leq 0.6$), capturing moderately uncertain fits. A value of 2 flags strongly discrepant measurements, which often indicate low SNR or continuum model degeneracies. The flag values \{0, 1, 2\} are roughly split into \{0.65,0.15,0.20\} for this criterion.
\item \texttt{balmer\_width\_flag}: compares the broad-line widths of \ha and \hb to assess cross-line consistency. A value of 0 corresponds to ($0 \leq \log \mathrm{FWHM}_{\mathrm{H}\beta} - \log \mathrm{FWHM}_{\mathrm{H}\alpha} < 0.25$), which is commonly observed in well-constrained fits. A value of 1 allows for small deviations ($-0.05 \leq \log \mathrm{FWHM}_{\mathrm{H}\beta} - \log \mathrm{FWHM}_{\mathrm{H}\alpha} < 0$). A value of 2 flags significant discrepancies between the two lines. These conditions are chosen based on the distribution in Figure~\ref{fig:balmer_width} and observed reverberation lag ratio for \ha and \hb in \citet{bentz_walsh_2010}. Large deviations, particularly in spectra with otherwise adequate SNR, are empirically associated with poorer-quality \hb fits and potential continuum-fitting degeneracies. The flag values \{0, 1, 2\} are roughly split into \{0.50, 0.15, 0.35\} for this criterion.
\item \texttt{super\_flag}: is the sum of the above five flags, with lower values meaning better spectrum decomposition and strong broad Balmer lines.
\item \texttt{spec\_flag}: indicates the basic quality of the spectrum based on visual inspection. The majority of spectra ($\sim$97 per cent) are labelled \texttt{good}, corresponding to cases where the data are reasonably well described by the fitted model. We emphasise that this flag reflects only the agreement between the data and the model, and does not guarantee that the decomposed components are physically meaningful; the additional flags described above should be used to identify reliable measurements.
Spectra labelled \texttt{bad\_all} or \texttt{bad\_blue} indicate that the entire spectrum, or the blue arm respectively, is of insufficient quality for reliable decomposition. In many such cases, alternative observations of the same target are available and yield acceptable results. 
All figures in this work that rely on spectral decomposition include a selection of \texttt{spec\_flag == good}.
\item \texttt{spurious\_hab\_flag}: identifies cases where the broad \ha component is likely non-physical. Because the narrow \ha and [N\,\textsc{ii}] lines are modelled with a single Gaussian component each, the broad \ha component can occasionally absorb residual flux from the wings of the narrow-line complex. Such cases are flagged when the broad component is both relatively narrow and weak compared to the narrow \ha line: $\mathrm{FWHM}_{\rm Br,H\alpha} < 10 \times \mathrm{FWHM}_{\rm Nr,H\alpha}$,
$L_{\rm Br,H\alpha} < 1.12 \times L_{\rm Nr,H\alpha}$,
and \texttt{hab\_snr} $> 1$.
For flagged objects, the broad \ha flux is redistributed among the narrow \ha and \nii components according to their fitted flux fractions. Approximately 4 per cent of the sample satisfies these criteria and is assigned \texttt{spurious\_hab\_flag} = 1; all other objects have a value of 0.
\item \texttt{spurious\_hbb\_flag}: identifies cases where one of the broad \hb components reaches the maximum allowed line width during fitting. In these objects, the broad \hb, Fe\,\textsc{ii}, and He\,\textsc{ii} components can become highly degenerate with the AGN and host-galaxy continua, producing an artificial broad feature extending across much of the \hb spectral region and leading to overestimated line luminosities. The flag takes three values: 0 indicates no issue; 1 indicates that the second broad \hb component reached the maximum allowed width and was removed; and 2 indicates that the primary broad \hb component was removed, following the approach of \citet{amrutha_masscorr_2026}. The luminosity of the removed component is added to the upper uncertainty of the retained broad \hb measurement. For all objects with \texttt{spurious\_hbb\_flag} $> 0$, Fe\,\textsc{ii} and He\,\textsc{ii} measurements are considered unreliable and are therefore excluded from the catalogue. 20 per cent of the sample have \texttt{spurious\_hbb\_flag} $> 0$.
\end{itemize}

\bsp	
\label{lastpage}
\end{document}